\pdfoutput=1
\documentclass[aps,prl, preprint]{revtex4-1}
\usepackage{amsmath}
\usepackage{amssymb}
\usepackage{fullpage}
\usepackage{graphicx}
\usepackage{xcolor}
\usepackage{afterpage}
\usepackage{mathtools}
\usepackage{color}
\usepackage{amsfonts}
\usepackage{xr}
\usepackage{eso-pic}
\makeatletter

\newcommand*{\BF}{Bi$_2$FeMnO$_6$}
\newcommand*{\FP}{First-principles}
\newcommand*{\fp}{first-principles}
\newcommand*{\Fe}{Fe$^{3+}$}
\newcommand*{\Mn}{Mn$^{3+}$}

\renewcommand{\fnum@figure}{\textbf{Figure \thefigure}} 
\makeatother
\begin{document}
\title{Coexistence of spin glass and ferroelectricity in highly ordered \BF{} epitaxial thin film}
\author{Lin Sun$^1$$^{\dagger}$}
\author{Yue-Wen Fang$^1$$^{\dagger}$}
\author{Jun He$^1$}
\author{Yuanyuan Zhang$^1$}
\author{Ruijuan Qi$^1$}
\author{Qing He$^2$}
\author{Rong Huang$^1$}
\author{Pinghua Xiang$^1$}
\author{Xiao-Dong Tang$^1$}
\author{Pingxiong Yang$^1$}
\author{Junhao Chu$^1$}
\author{Ying-Hao Chu$^3$}
\author{Chun-Gang Duan$^{1}$}
\thanks{E-mail: wxbdcg@gmail.com or cgduan@clpm.ecnu.edu.cn}

\affiliation{$^1$Key Laboratory of Polar Materials and Devices, Ministry of Education, Department of Electronic Engineering, East China Normal University, Shanghai 200241, China}

\affiliation{$^2$Department of Physics, Durham University, Durham DH1 3LE, UK}

\affiliation{$^3$Department of Materials Science and Engineering, National Chiao Tung University, Hsinchu, 30010, Taiwan, ROC}


\date{\today}
\begin{abstract}
Highly ordered \BF{} epitaxial thin films have been successfully grown on SrTiO$_3$ substrate. Both  high-flux synchrotron X-ray diffraction reciprocal space mapping and high resolution transmission electron microscopy confirmed the alternative alignment of Fe and Mn along [111] direction of \BF{} films. Magnetic and ferroelectric properties of \BF{} films are characterized and analyzed. The room-temperature ferroelectricity is well kept in \BF{} film as expected. However, it is very interesting that \BF{} film exhibits a typical spin-glass behavior and very weak magnetism rather than a ferri$\slash$ferromagnetism as generally believed. Our \fp{} calculations suggest a spin frustration model for \BF, which can well explain the intriguing magnetic property of \BF{} film.
\end{abstract}

\maketitle

\section*{Introduction}
Multiferroic materials have attracted significant attention in recent years because of their intriguing physical phenomena and huge potential applications in multifunctional devices, such as sensors and memory devices for storage data\cite{Bibes.Nat.Mat.2008,Cheong.NatMat.2007,Ramesh.NatMat.2007}{}. Single-phase multiferroics are very rare because ferroelectricity generally requires empty $d$-orbitals whereas ferromagnetism needs partial filled $d$-orbitals, which is commonly incompatible\cite{Spaldin.JPCB.2000}{}. As exceptions, bismuth based perovskite oxides BiFeO$_3$ (BFO) and BiMnO$_3$ (BMO) are well-known multiferroic materials. The coexistence of ferroelectricity and (anti)ferromagnetism in these compounds is thought to be realized by the combination of magnetic element (Fe or Mn) at B site and Bi 6$s^2$ lone-pair electrons at A site. However, BFO is a canted antiferromagnet (N\'{e}el temperature T$_N$ $\sim$ 640 K), which leads to very weak ferromagnetism, although it has a very high ferroelectric Curie temperature (T$_c$ $\sim$ 1103 K)\cite{Smolenskii.ferro.1982}{}. As for BMO, its ferroelectric Curie temperature approaches to 450 K, whereas it has a low ferromagnetic (FM) Curie temperature ($\sim$ 105 K) far below room-temperature (RT)\cite{Darrell.APL.2004}{}. Moreover, BMO is metastable, which makes its fabrication very difficult. Previous studies have pointed out that the bulk phases of BMO have to be synthesized by the high-pressure method\cite{Faqir.JSSC.1999} and the growth of pure phase BMO epitaxial films can only be achieved by strain engineering\cite{Darrell.APL.2004,Lee.APL.2010,Luca.APL.2013}{}. These drawbacks of BFO and BMO materials inevitably limit their practical applications on spintronic devices, especially in RT environment. \\

To overcome these aforementioned drawbacks, double-perovskite \BF{} (BFMO) was recently proposed as a good candidate for RT single-phase multiferroic material\cite{BiLei.PRB.2008,Zhao.APL.2009}{}. In BFMO, Fe and Mn ions were reported to energetically favor high spin state, i.e., Fe$^{3+}$ and Mn$^{3+}$\cite{BiLei.PRB.2008}{}, and the coupling of dissimilar ions of Fe$^{3+}$($d^5$)-O-Mn$^{3+}$($d^4$) were suggested to be antiferromagnetic (AFM) in previous studies\cite{BiLei.PRB.2008,DuYi.APL.2010,Choi.AFM.2014}{}. This antiparallel coupling is believed to provide a net magnetization as Fe$^{3+}$ and Mn$^{3+}$ ions have unequal magnetic moments. Thus, we expect ferrimagnetism could be induced via the alternative periodic arrangement of Fe$^{3+}$ and Mn$^{3+}$ at B sites along [111] direction of BFMO. Similar idea has been proposed by Rabe \emph{et al.} \cite{Rabe.PRB.2010,Rabe.PRL.2010} to predict the coexistence of ferroelectricity and ferrimagnism in BFO-BMO checkerboard (i.e., the (110)-oriented BFO-BMO superlattice). Many attempts therefore have been made to fabricate BFMO and explore their multiferroic properties\cite{BiLei.PRB.2008,Zhao.APL.2009,Kundu.APL.2008,Mandal.PRB.2010,Delmonte.PRB.2013,Zhao.JAP.2010,DuYi.APL.2010,Zhao.2010,Cortie.APL.2012,Liu.JAP.2013,Choi.APL.2011,Choi.AFM.2014,Miao.APL.2011,ChenAiping.AM.2013}{}. However, experimentally it is hard to achieve the exactly ordered BFMO supercell structure. At present, only disordered BFMO bulk and films, in which Fe and Mn ions are randomly distributed at B sites, have been synthesized, and no convincing experimental evidence for Fe-Mn ordering along [111] direction is found\cite{Kundu.APL.2008,Mandal.PRB.2010,Delmonte.PRB.2013,Zhao.JAP.2010,Choi.AFM.2014}{}. In addition, recent neutron diffraction experiment further confirmed Fe and Mn disordering at B sites at RT in BFMO epitaxial films\cite{Cortie.APL.2012}{}.   \\

Based on first-principles calculations, BFMO is a metastable compound in which Fe-Mn ordering is not favored\cite{BiLei.PRB.2008}{}. Note that Nechache \emph{et al}. have made systematical studies on the fabrication and characterization of double perovskite Bi$_2$FeCrO$_6$ (BFCO) epitaxial films\cite{Nechache.JAP.2009,Nechache.JPCM.2012,Nechache.APL.2006,Nechache.Nat.Phot.2015}{}. They not only grew metastable BFCO films by pulsed laser deposition (PLD) through strain engineering, but also confirmed Fe-Cr ordering along [111] direction through asymmetric X-ray diffraction (XRD) method. Fe-Cr ordering indeed gives rise to RT ferrimagnetic behavior, and consequently BFCO films exhibit a good RT multiferroic properties\cite{Nechache.JAP.2009,Nechache.JPCM.2012}{}. In addition, Bi$_2$NiMnO$_6$ epitaxial films with ordered double-perovskite structure have been grown on (001) SrTiO$_3$ (STO) substrates by PLD and exhibit multiferroic properties at 100 K\cite{Sakai.APL.2007}{}. Fe-Mn ordering was also achieved in LaFe$_{0.5}$Mn$_{0.5}$O$_3$ film grown on STO (111) substrates\cite{Ueda.APL.2001,Ueda.PRB.1999}{}. Therefore, these pioneering works suggest that it is feasible to grow epitaxial BFMO films with Fe-Mn ordering. Furthermore, Mn$^{3+}$ is a Jahn-Teller ion, which often induces exotic phenomena in perovskite oxides, consequently we anticipate BFMO would demonstrate more interesting properties. Finally, Fe-Mn ordering is expected to reduce the band gap to 1$\sim$2 eV\cite{Xu.APL.2010}{}, thus BFMO could be a promising candidate for ferroelectric photovoltaic materials. These above reasons motivate us to study BFMO.\\

In this work, we have successfully fabricated double-perovskite BFMO epitaxial films on STO (001) substrates by PLD, and further confirmed the Fe-Mn ordering along [111] direction using synchrotron reciprocal space mapping (RSM), asymmetric XRD and high resolution transmission electron microscopy (HRTEM) measurements. RT ferroelectricity is revealed in these BFMO epitaxial films by the ferroelectric measurement. The interesting part is, contrary to what we expected, the characterization and analysis of magnetism indicate that [111] Fe-Mn ordering of BFMO does not result in RT ferri$\slash$ferromagnetism. Instead, BFMO only shows a very weak magnetism due to a special magnetic state--spin glass. Thus, the magnetic structure of BFMO is completely different from that of BFO. Our research confirms that the metastable BFMO film with [111] Fe-Mn ordering can be realized experimentally, and it favors a spin frustrated magnetic structure.

\section*{Results}
\textbf{Structure characterization and analysis.} The BFMO films with different thickness were grown on [001]-oriented STO substrates by PLD. Both 40 and 190 nm-thick BFMO films are pure phase since all diffraction peaks can be indexed, as shown in Supplementary Fig. S1a. Only (00$l$) peaks ($l$ = 1, 2, 3, 4) of BFMO films and the corresponding peaks of STO substrates are observed in Fig. S1b, which indicates that BFMO films are well oriented along the normal direction of STO substrate surface. For 40 nm BFMO film, the fringe-peak oscillation is clearly observed in 2theta-Omega scan in the vicinity of (001) reflection (see Supplementary Fig. S1c). The full width at half maximum (FWHM) values of the rocking curve scan around (002) diffraction peak are 0.05$^\circ$ for BFMO and 0.03$^\circ$ for STO (see Supplementary Fig. S1d). The root-mean-squared roughness of 40 nm BFMO (measured area $2\times 2\ \mu \rm{m^2}$ measured by atomic force microscopy is about 1.39 nm (see Supplementary Fig. S1e). These observations indicate that 40 nm BFMO film deposited on STO (001) substrate is a smooth and high-quality epitaxial film. BFMO film grown on STO substrate suffers from a compressive strain and when the thickness of BFMO film increases up to 190 nm, it inevitably becomes relaxed, at least partially. The broader (00l) peaks at 2theta-Omega scan and high FWHM ($\sim$ 0.24$^\circ$) of (002) peak at rocking curve for 190 nm BFMO film reflect this fact indirectly (see Supplementary Fig. S1b-d). The image of 190 nm BFMO film shows it has a smooth surface in rectangular grains of epitaxial films like 40 nm BFMO (see Supplementary Fig. S1f).\\

\begin{figure}[!ht]
\centering
\includegraphics[width=30pc,clip]{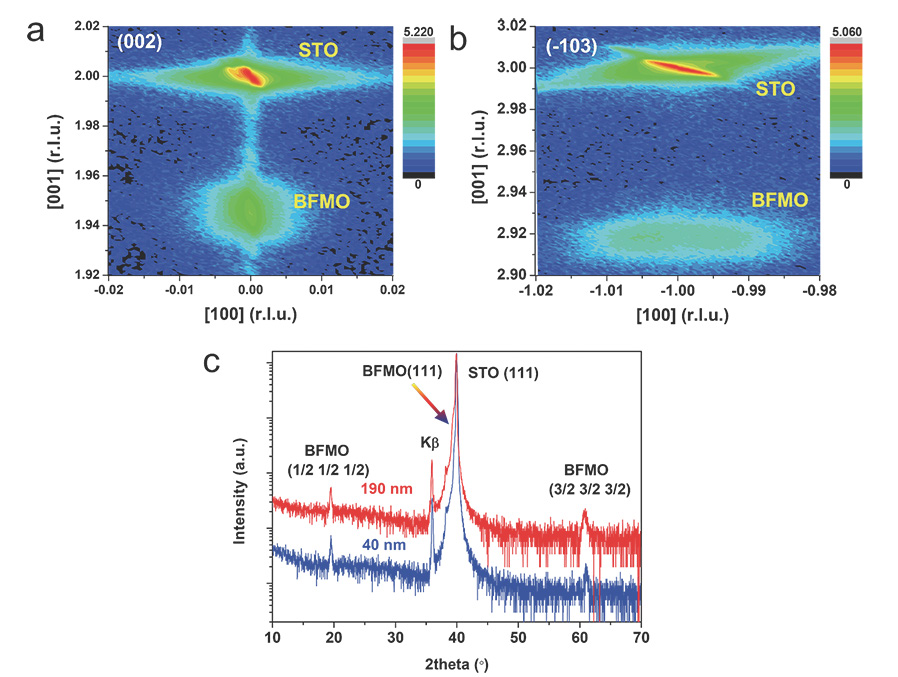}
\caption{\textbf{Structural characterization of BFMO films.} (a) RSM of (002) reflection of 40 nm-thick BFMO film and STO substrate. (b) RSM of (-103) reflection of 40 nm-thick BFMO film and STO substrate. (c) Asymmetric (111) 2theta-Omega XRD scan of 40 nm-thick and 190 nm-thick BFMO films grown on (001)-oriented STO substrates, $K_{\beta}$ symbol represents the Cu-$\rm K\rm _{\beta}$ signal from STO substrate.} \label{fig:FIG1RSM}
\end{figure}

To further analyze the epitaxial strain state of BFMO films, RSM was used to record around symmetric (002) and asymmetric (-103) pseudo-cubic (${pc}$) reflections. Fig. \ref{fig:FIG1RSM}a,b show RSM patterns for 40 nm BFMO film, measured in the vicinity of (002) and (-103) reflections, respectively. The RSM patterns are plotted by reciprocal lattice units (r.l.u.) of STO substrate (1 r.l.u. = 2$\pi \slash 3.905$ \AA). RSM around (002) reflection clearly demonstrates the full orientation of BFMO film along the perpendicular direction to the surface of STO (001) substrate, and $c_{pc}$ is determined to be $\sim 4.015$ \AA\ according to the vertical distance between the reciprocal points of BFMO and that of STO. The fringe peak oscillation is also appeared in the vicinity of (002) reciprocal point along [001] direction. It is caused by the interference of thin film thickness and generally only exists in very thin films ($<$ 100 nm or less). RSM around (-103) reflection exhibits that the reciprocal points of BFMO and STO are the same position along the horizontal axis (i.e., in-plane [-100] direction), revealing the same in-plane lattice parameters ($a_{pc}$\ $\sim 3.905$ \AA) of BFMO film and STO substrate. This observation confirms the high-quality epitaxial growth of BFMO film on STO (001) substrate, and the in-plane strain caused by lattice mismatch between BFMO and STO is completely kept in BFMO film. This residual strain is compressive and thus the out-of-plane lattice parameter $c_{pc}$ ($\sim 4.015$ \AA) is longer than in-plane lattice parameter $a_{pc}$ ($\sim 3.905$ \AA) in 40 nm BFMO film. For 190 nm BFMO film, RSM of (002) and (103) reflections demonstrate that the reciprocal space points of BFMO are still located at the same position as STO substrate along horizon axis (i.e., [100] direction), as can be seen in Supplementary Fig. S2. However, it is obvious that the reciprocal space points of 190 nm BFMO film are partially dispersed or relaxed points, in contrast with those of 40 nm BFMO film which are small condensed points. Therefore, 190 nm BFMO film is also epitaxially grown on STO (001) substrate but in partially relaxed state.\\

Though our BFMO films are deposited on [001]-oriented STO substrates, the ordering of Fe and Mn is along [111] direction (i.e., BFO-BMO superlattice along [111]). Thus, the asymmetric XRD measurement around (111) reflection is a simple and feasible way to detect the existence of Fe and Mn ordering along [111] direction. If there are some ($m$/2, $m$/2, $m$/2) ($m$ = 1, 3 ...) superlattice peaks in the asymmetric (111) scan besides (111) peaks, Fe and Mn [111] ordering can be primitively confirmed in BFMO films. The asymmetric scan results of BFMO films on STO (001) substrates are shown in Fig. \ref{fig:FIG1RSM}c. As we expected, two superlattice peaks clearly appear at 19.5$^\circ$ and 61.0$^\circ$, besides (111) main peaks of BFMO and STO. The presence of these two superlattice peaks is due to the Fe and Mn alternative arrangement along [111] direction, and they are indexed by (1/2, 1/2, 1/2) and (3/2, 3/2, 3/2) superlattice reflections, respectively. According to Bragg's law, the interplaner spacing $d$(1/2, 1/2, 1/2) and $d$(1, 1, 1) of BFMO films are equal to 4.55 and $2.28 $ \AA, respectively. This means that the periodicity of (1/2, 1/2, 1/2) superlattice reflection is $4.55 $ \AA, which is almost twice of that of (111) reflection (i.e., $d$(1/2,1/2,1/2) $\approx$ 2$d$(1,1,1)).\\

\begin{figure}[!ht]
\centering
\includegraphics[width=32pc,clip]{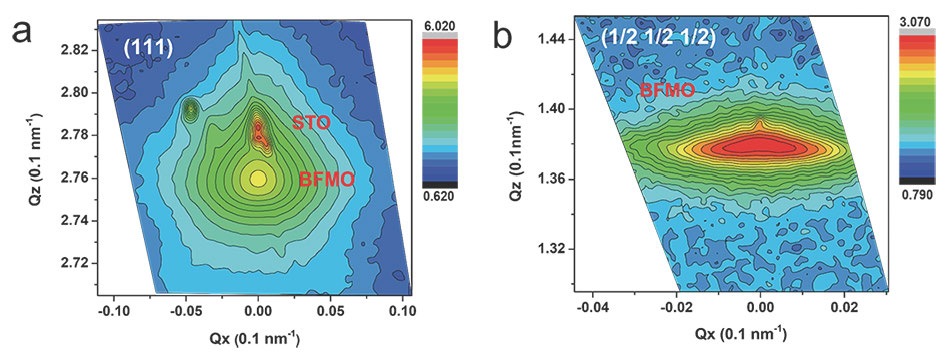}
\caption{\textbf{Direct identification of Fe and Mn ordering along [111] direction using synchrotron X-ray diffraction.} (a) RSM of (111) reflection of 190 nm-thick BFMO film and STO substrate. (b) RSM of superlattice (1/2, 1/2, 1/2) reflection of 190 nm-thick BFMO film. } \label{fig:FIG2XRD}
\end{figure}

It should be emphasized that Fe and Mn have very similar scattering factor and it is quite difficult to detect very weak diffraction peak from Fe and Mn [111] ordering by the conventional XRD measurement, especially for RSM. Thus, to provide more convincing evidence of Fe-Mn ordering in BFMO films, high-flux synchrotron X-ray diffraction measurement at Shanghai Synchrotron Radiation Facility was employed. The results, i.e., RSM about (111) and (1/2 1/2 1/2) reflections for 190 nm BFMO film are shown in Fig. \ref{fig:FIG2XRD}a, b, where the reciprocal space coordinates is \AA$^{-1}$. A sharp reciprocal point (111) originating from STO and the relatively relaxed reciprocal point (111) coming from BFMO are presented in RSM (111) image. Their reciprocal space length Q$_Z$ corresponding to interplanar spacing $d$(111) for STO and BFMO are 2.78 and 2.76 \AA$^{-1}$, respectively. It is well known that $d$(111)= 2$\pi$/Q$_Z$, thus $d$(111) for STO and BFMO are 2.26 and 2.28 \AA, respectively. More importantly, RSM (1/2, 1/2, 1/2) directly gives the apparent reciprocal point (1/2, 1/2, 1/2) originating from Fe and Mn ordering in BFMO film. Its reciprocal space length is about 1.38 \AA$^{-1}$, which suggests that the interplanar spacing $d$(1/2 1/2 1/2) = 2$\pi$/(1.38 \AA$^{-1}$) $\approx$ 4.55 \AA. For BFMO, $d$(1/2 1/2 1/2) $\approx$ 2$d$(111) is confirmed by RSM, which is consistent with asymmetric XRD data in Fig. \ref{fig:FIG1RSM}c. RSM (111) and (1/2 1/2 1/2) patterns for 40 nm BFMO are in well agreement with that of 190 nm BFMO (supplementary Fig. S3). Hence, by means of synchrotron RSM (111) and (1/2, 1/2, 1/2) measurement, we unambiguously identify the Fe and Mn ordering in our BFMO films, i.e., Fe and Mn ions arrange alternatively along [111] direction and form the rock-salt-type double-perovskite BFO-BMO superlattice.\\

\begin{figure}[!ht]
\centering
\includegraphics[width=28pc,clip]{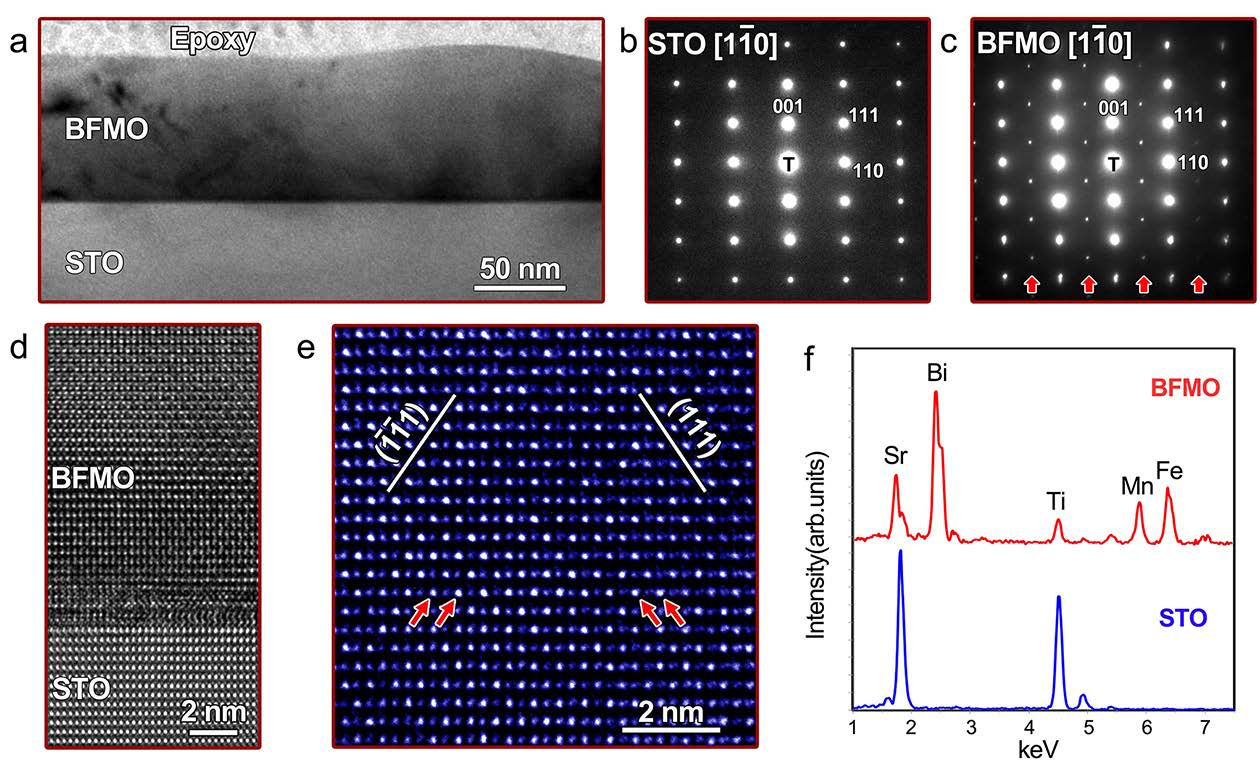}
\caption{\textbf{Confirmation of the ordering of B sites in the BFMO thin film by HRTEM. (a) Typical bright field TEM image of the thin film.} (b) SAED patterns of the STO substrate along the [1$\bar{1}$0] zone axis. (c) SAED patterns of the BFMO thin film along the [1$\bar{1}$0] zone axis showing the superlattice reflections in the middle of the transmission spot and the 111 diffraction spots. (d) HRTEM image of the BFMO/STO interface. (e) Typical HRTEM image of the BFMO thin film revealing the ordering of the (111) and [$\bar{1}\bar{1}$1] planes. (f) Typical EDS spectra of the BFMO thin film and the STO substrate.} \label{fig:FIG3TEM}
\end{figure}

Synchrotron X-ray RSM (1/2, 1/2, 1/2) definitely provides a strong support of the Fe and Mn ordering along [111] direction in BFMO films. In order to obtain more detailed microstructural information of the BFMO thin film and confirm the ordering of Fe and Mn at atomic scale, a cross-sectional thin foil specimen was prepared by the conventional methods and then observed by HRTEM. Figure \ref{fig:FIG3TEM}a shows a typical bright filed image of the BFMO thin film. It can be seen that the contrast of BFMO thin film is very uniform and no grain boundaries can be observed, indicating a single crystal BFMO thin film was obtained. The thin film has a very flat BFMO/STO interface. To reveal the cation ordering at B sites along the [111] direction, [1$\bar{1}$0] zone axis was selected for HRTEM observations. Figure \ref{fig:FIG3TEM}b,c exhibit the selected area electron diffraction (SAED) patterns of the STO substrate and the BFMO thin film along [1$\bar{1}$0] the zone axis, respectively. It reveals a cubic-on-cubic epitaxial relationship between the BFMO thin film and STO substrate. Interestingly, the superlattice reflections (marked by arrows) appearing in the middle of the transmission spot and the (111) diffraction spots clearly indicate the presence of the ordering along the [111] direction in the BFMO thin film, as shown in Fig. \ref{fig:FIG3TEM}c, which is consistent with the XRD results. Fig. \ref{fig:FIG3TEM}d gives a HRTEM image of the BFMO/STO interface, which clearly shows the epitaxial growth of the BFMO on the STO substrate with the atomic level sharp interface, indicating high quality of the sample. More importantly, we could observe the long-range ordering of the (111) plane along the [111] direction with atomic resolution, as indicated by the arrows in Fig. \ref{fig:FIG3TEM}e. Both the (111) and [$\bar{1}\bar{1}$1] planes demonstrate bright and dark contrast alternately, revealing a twofold superstructure of the (111) plane in the pervoskite structure. Furthermore, the X-ray energy dispersive spectroscopy (EDS) analyses revealed that the Bi:Fe:Mn cation ratio is nearly 2:1:1 in the thin film, as shown in Fig. \ref{fig:FIG3TEM}f, consistent with the designed composition. Thus, the combination of HRTEM observations and EDS analyses unambiguously confirmed the ordering of Fe and Mn ions in the (111) plane along the [111] direction in the BFMO thin film.\\

\begin{figure}[!ht]
\centering
\includegraphics[width=28pc,clip]{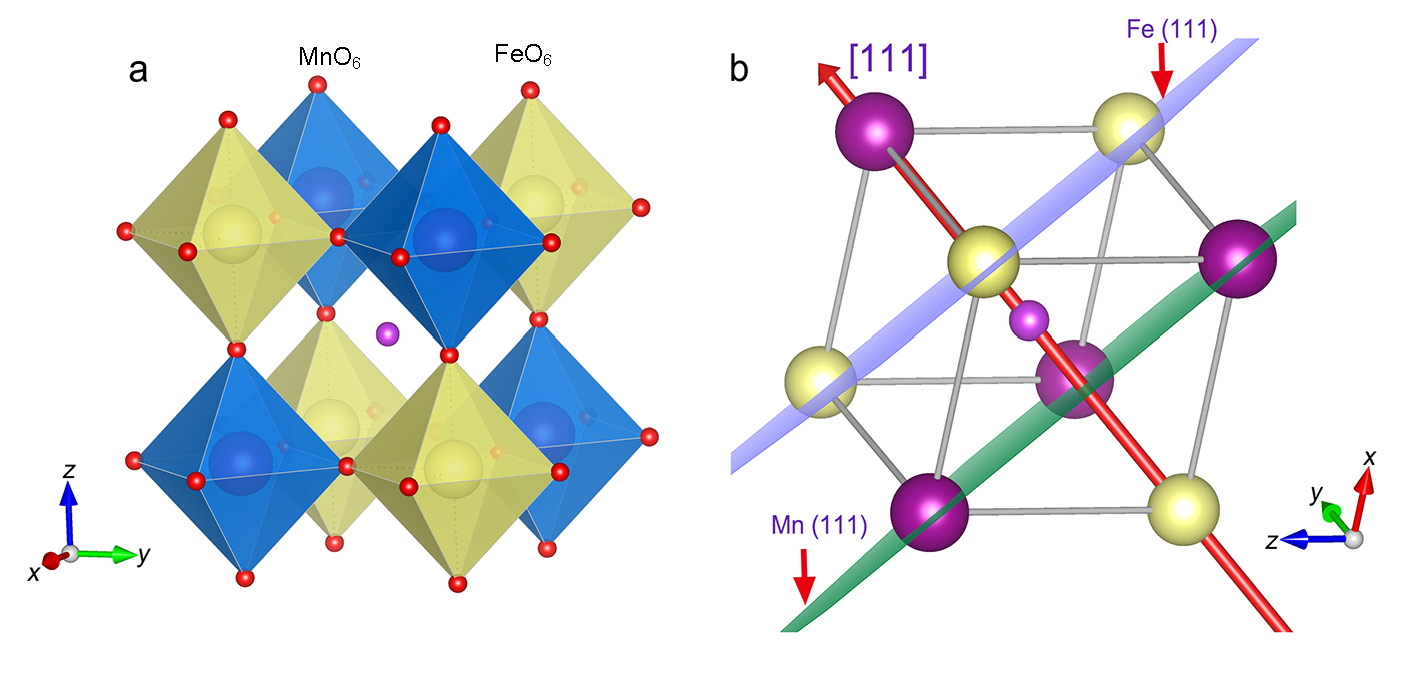}
\caption{\textbf{Schematic representation of crystal structure of rock-salt-structured double-perovskite BFMO.} (a) The crystal structure of BFMO grown on STO (001) substrate. (b) The B-site ordering along [111] direction. Red arrow denotes the [111] direction.} \label{fig:FIG4struct}
\end{figure}

Based on diffraction techniques, we can conclude that Fe and Mn ordering along [111] direction exists in  the deposited BFMO films on STO (001) substrates. Fig. \ref{fig:FIG4struct}a illustrates the crystal structure of double-perovskite BFMO epitaxially grown on STO (001) substrate. As shown in Fig. \ref{fig:FIG4struct}b, Fe and Mn ordering in BFMO adopts the rock-salt configuration and thus BFMO can also be referred to [111]-oriented BFO-BMO superlattice. As BFO is a well-known ferroelectric and canted AFM material\cite{Ederer.PRB.2005,Lebeugle.PRL.2008}{}, while double-perovskite BFMO has similar crystal structure as BFO except for the Fe-Mn ordering along [111] direction. A question naturally arises: what is the magnetic structure of BFMO?

\begin{figure}[!ht]
\centering
\includegraphics[width=24pc,clip]{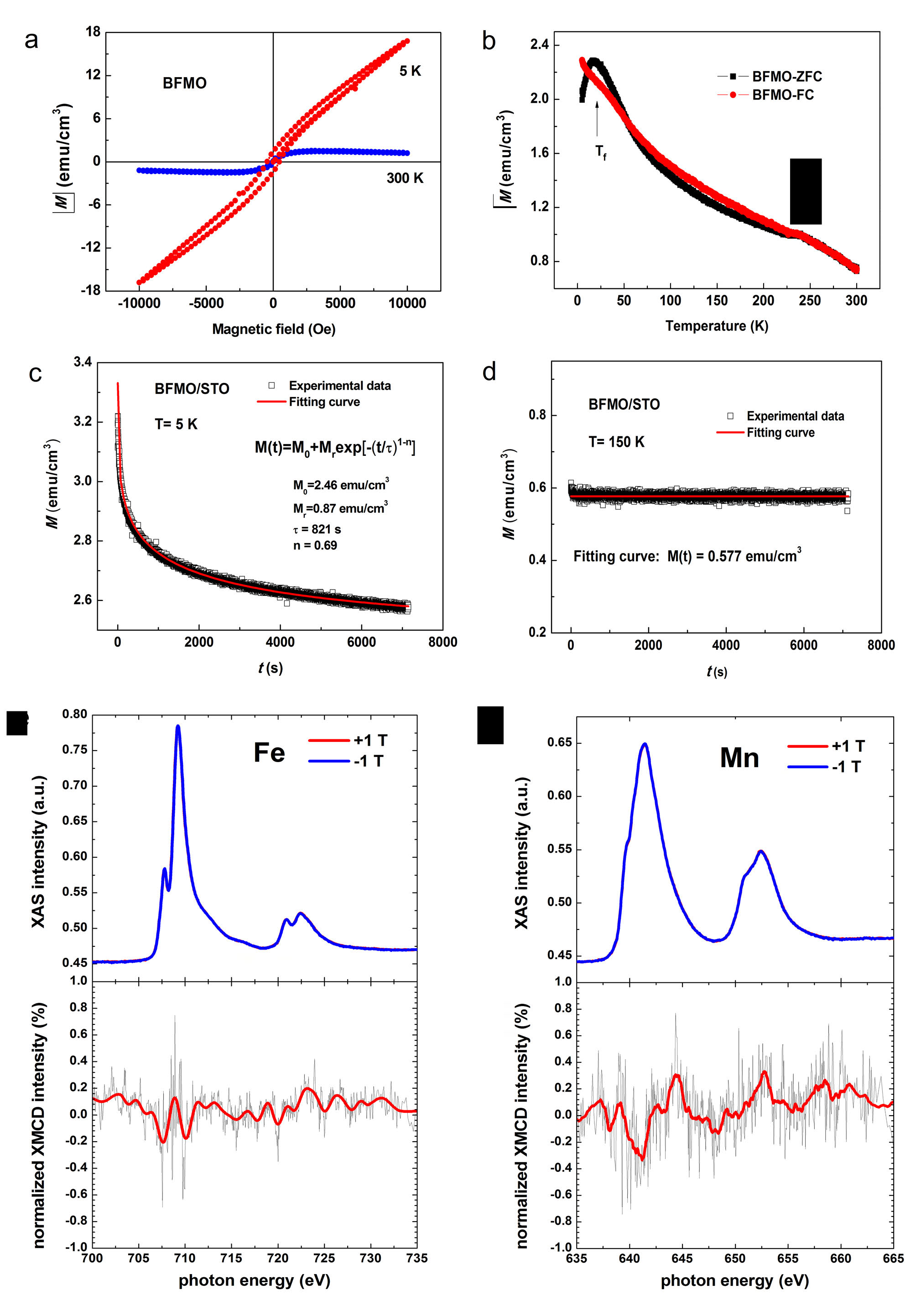}
\caption{\textbf{Magnetic characterization of BFMO film.} In-plane magnetic hysteresis ($M$-$H$) curves of BFMO film epitaxially grown on STO substrate after the subtraction of the diamagnetic signal of STO substrate, (a) measured at 5 K and 300 K; (b)  temperature dependent magnetization($M$-$T$) curves of BFMO film (5 K $\sim$ 300 K);  the measured and fitted time dependent remnant magnetization at 5 K (c) and 150 K (d) for BFMO film; XAS and XMCD spectra for Fe (e) and Mn (f) in BFMO film.} \label{fig:FIG5mag}
\end{figure}

\textbf{Magnetism characterization and analysis.} Magnetic properties of 190 nm BFMO film were measured by superconducting quantum interference device (SQUID) and the magnetic field was applied along the in-plane direction, parallel to the [100] direction of the STO substrate. Fig. \ref{fig:FIG5mag}a shows the magnetic hysteresis ($M$-$H$) loop of BFMO film at 5 K and 300 K, respectively. The saturated magnetization ($M_s$) value of BFMO film at 300 K are about 1.8 emu/cm$^3$, which corresponds to $\sim 0.02$ $\rm \mu_B$/Fe-Mn pair. The in-plane magnetic hysteresis ($M$-$H$) loop of 40 nm BFMO film at 300 K and 10 K can be seen in Supplementary Fig. S4. $M_s$ of 40 nm BFMO film at 10 K is about 0.2 $\rm \mu_B$/Fe-Mn pair. The $M_s$ discrepancy between 40 and 190 nm BFMO films is likely to be related with epitaxial strain\cite{Choi.APL.2011}{}. The magnetic behavior of 190 nm BFMO should have more intrinsic character because the effect of epitaxial strain can be neglected. Hence, the $M$-$H$ curve at 300 K indicates that BFMO film has very weak magnetism. It is widely accepted that BFO is a canted antiferromagnet and thus only has very small magnetic moment ($\sim 0.02$ $\rm \mu_B$/Fe)\cite{Ederer.PRB.2005,Albrecht.PRB.2010}{}. But for BFMO with [111] Fe-Mn ordering, X-ray photoelectron spectroscopy (XPS) analysis in Supplementary Fig. S5 has verified that both Fe and Mn ions are mainly in trivalent (+3) state. It is reported that both Fe$^{3+}$ and Mn$^{3+}$ in BFMO have high-spin (HS) state electron configuration ($t_{2g}^{3}$$e_{g}^{2}$ for Fe$^{3+}$ and $t_{2g}^{3}$$e_{g}^{1}$ for Mn$^{3+}$)\cite{BiLei.PRB.2008,DuYi.APL.2010}{}. Since the magnetic moments of Fe and Mn are different (5 $\rm \mu_B$ for \Fe and 4 $\rm \mu_B$ for \Mn), AFM structure of BFMO should result in a net magnetic moment of 1 $\rm \mu_B$ per Fe-Mn pair, and thus the ferrimagnetism is expected. As a result, the macroscopic magnetic moment of BFMO film is about 1 $\rm \mu_B$ per Fe-Mn pair if the AFM structure model of BFO is valid for BFMO. However, the magnetic moment of BFMO is only 0.02 $\rm \mu_B$/Fe-Mn pair, which suggests the magnetic structure of BFMO is different from that of BFO.

More interestingly, it can be observed that $M$-$H$ loop at 5 K, which is entirely different from that of 300 K, does not exhibit the saturation in magnetic field up to 1 T. The out-of-plane zero-field-cooled (ZFC) and field-cooled (FC) temperature dependent magnetization ($M$-$T$) curves for BFMO film were measured under 1000 Oe from 5 K to 300 K, which is shown in Fig 5b. The $M$-$T$ curve of STO substrate was measured and subtracted from BFMO/STO data. The ZFC and FC magnetization values of BFMO increase with decreasing the measured temperature in the range of 50 K-300 K. Below 50 K, ZFC and FC magnetization curve begin to split and a peak appears at around 18 K in ZFC curve whereas FC curve continues to increase with decreasing the temperature up to 5 K. This phenomenon exhibits a signature of spin-glass behavior, since the evidences of spin-glass state in oxide include that there is a difference between ZFC and FC curves at low temperature, and a cusp (or peak) exists in ZFC curve\cite{Scott.AM.2009}{}. Similar observation has been reported in LaFe$_{0.5}$Mn$_{0.5}$O$_3$ film and La doped BFMO ceramics with spin-glass state\cite{Giri.JAP.2006}{}. Thus, BFMO is likely to be a spin glass and the spin-freezing temperature (T$_f$) is about 18 K, corresponding to the peak temperature in ZFC curve. It can be understood that the $M$-$H$ loop at 5 K reveals weak ferromagnetism in BFMO film due to spin freezing and it does not show the saturation even at 1 T field like other spin glass systems\cite{Singh.PRB.2008}{}.\\

It should be pointed out that the deviation of ZFC and FC curves and the existence of peak in ZFC at low temperature region are not sufficient evidence for spin glass behavior. For example, this behavior can also appear in superparamagnetic and AFM materials. However, spin glass state is a metastable phase and the phenomenon of aging is one of important characteristics of spin glass. Both superparamagnetic and AFM systems have no aging behavior. To further confirm the spin glass behavior of BFMO film, we have measured the relaxation of magnetization at 5 K ($<$ T$_f$) and 150 K ($>$ T$_f$). The BFMO sample was cooled in 10000 Oe field from 300 K to 5 K or 150 K, waited for 30 s, and then decreased the field to 500 Oe and measured the time-dependent magnetization. An extended exponential decay can be applied to describe the magnetic relaxation of spin glass, which is as follows:
\begin{equation}\label{EquAging}
  M(t)= M_0 + M_{r}exp\left[-{(\frac{t}{ \tau })}^{1-n}\right]  \ ,
\end{equation}
where $M_0$ stands for the nonrelaxed magnetization under 500 Oe field, $M_r$ contributes to the observed relaxed effects, the time constant $\tau$ and exponent $n$ reflect the relaxation rate of spin glass. If $0  < n < 1$, it represents spin glass system\cite{Xu.Sci.rep.2015,Diep.FCC.2014,Cardoso.PRB.2003}{}. Fig. \ref{fig:FIG5mag}c shows the time dependent remnant magnetization at 5 K for BFMO film and there is a typical relaxation behavior measured at 5 K for BFMO film. Red solid curve is the best fitting curve to the measured data with equation (1) and the fitting parameters are also listed in Fig. \ref{fig:FIG5mag}c. The fitting to experimental data is excellent. The $n$ and $\tau$ parameter of BFMO is 0.69 and 821 s, respectively. These two parameters are close to those of BFO/BMO, which is a spin glass system\cite{Xu.Sci.rep.2015}{}. In comparison, Fig. \ref{fig:FIG5mag}d gives the time dependent remnant magnetization at 150 K ( $>$ freezing temperature T$_f$) for BFMO film. No relaxation is observed at 150 K and fitting to measured data also verifies it. Hence, the combination of $M$-$T$ curve (Fig. \ref{fig:FIG5mag}b) and $M$-$t$ relaxation (Fig. \ref{fig:FIG5mag}c,d) eventually identify that BFMO has a typical spin-glass behavior.\\

Why does BFMO exhibit a spin-glass-like behavior? Site disorder and spin frustration generally give rise to spin-glass state. Our BFMO samples, however, have been proved to be highly ordered single crystalline film. To further explore the magnetic structure of BFMO, X-ray magnetic circular dichroism (XMCD) measurement, a powerful element-specific technique to investigate the magnetic response of materials\cite{Frank.CheRev.2001,HeQing.NatComm.2011} was adopted. Magnetic moments of Fe and Mn elements can be characterized or estimated respectively by XMCD. Fig. 5e,f show X-ray absorption spectroscopy (XAS) at $L$-edges and XMCD of Fe and Mn element of BFMO film, respectively. XMCD is defined as the difference between the XAS with the polarization of the incident X-ray to be parallel and anti-parallel to the applied magnetic field (1 T in the measurements here). And its intensity represents the element-specific net magnetization. As shown in Fig. \ref{fig:FIG5mag}c,d, +3 valence states for both Fe and Mn are confirmed by the XAS across Fe and Mn $L$-edges. However, no measurable XMCD (signal below noise level) can be observed from Fe or Mn $L$-edges of BFMO. This experimental result is very interesting. It implies that both (111) Fe plane and (111) Mn plane demonstrate no or very weak magnetism. XMCD measurement further suggests that the magnetic model of BFO is not applicable to BFMO, and there should be a different magnetic structure for BFMO. Based on SQUID and XMCD experimental data, we argue that a spin frustration magnetic structure occurs in (111) Fe or Mn plane for BFMO. The presence of spin frustration will lead to no or very weak magnetism, which can be used to explain the magnetic behavior of BFMO. \\

\begin{figure}[!ht]
\centering
\includegraphics[width=24pc,clip]{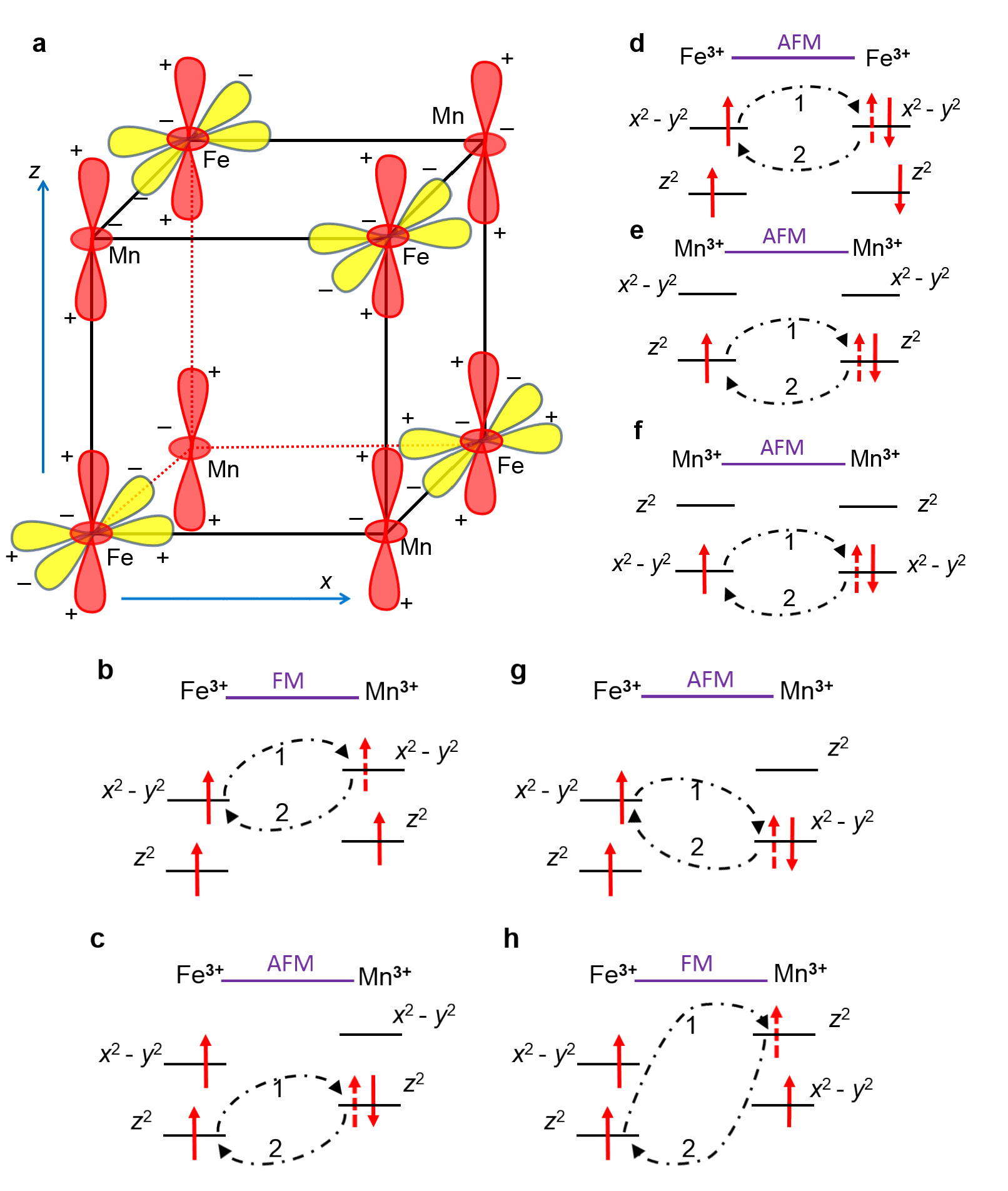}
\caption{\textbf{Active orbitals and virtual hoping process.} (a) The respective active oritals of \Fe\ and \Mn\ are plotted. (b) The in-plane NN FM coupling of \Fe-\Mn contributed by the virtual hopping from a half-filled $d_{x^2-y^2}$ orbital to an empty $d_{x^2-y^2}$ orbital. (c) The out-of-plane NN AFM coupling of \Fe-\Mn{} contributed by the virtual hopping between two half-filled $d_{z^2}$ orbitals. (d) and (e) describe the virtual electron hopping processes of NNN \Fe-\Fe\ and \Mn-\Mn\ superexchange, respectively. (f) describes the virtual electron hopping process of \Mn-\Mn\ after the reversal of energy level of \Mn. Whatever the NN \Fe-\Mn\ superexchange is, NNN \Fe-\Fe\ and \Mn-\Mn\ superexchange favors AFM coupling owning to the virtual electron hoping between two half-filled orbitals. Due to the reversal of energy level of $d_{x^2-y^2}$ and $d_{z^2}$ of \Mn, in-plane \Fe-\Mn\ superexchange becomes antiferromagnetically coupled between two half-filled $d_{x^2-y^2}$ orbitals(g), and the out-of-plane \Fe-\Mn\ becomes ferromagnetically coupled contributed by the virtual hopping from a half-filled $d_{z^2}$ orbital to an empty $d_{z^2}$ orbital(h).} \label{fig:FIG6orbitals}
\end{figure}

\textbf{Orbital model analysis and simulations of magnetic structures.} As mentioned before, previous studies\cite{BiLei.PRB.2008,DuYi.APL.2010,Choi.AFM.2014} suggested that \Fe-\Mn\ coupling should be AFM, consequently BFMO should produce sizable net magnetism and no frustration can be created. Nevertheless, all magnetic measurements suggest that BFMO sample is a spin frustrated system featuring very weak or no magnetism. This pushes us to reconsider the coupling between \Fe\ and \Mn. In general, an octahedral crystal field in cubic lattice leads to double orbital degeneracy of $e_g$, however, in our case with Jahn-Teller ions \Mn, the orbital degeneracy of $e_g$  is lifted by reduction of symmetry of lattice where $a_{pc}$ $\approx 3.905$ \AA\ and $c_{pc}$ $\approx 4.015$ \AA, thus $e_g$ orbital splits into $d_{x^2-y^2}$ and $d_{z^2}$ orbitals whose shapes are described in Fig. \ref{fig:FIG6orbitals}a. In paraelectric phase of BFMO, the regular octahedral crystal fields push up the $d_{x^2-y^2}$ level slightly higher than $d_{z^2}$ level for magnetic ions, as shown in Fig. \ref{fig:FIG6orbitals}b,c,d,e. In this case, the in-plane \Fe\ and \Mn\ are ferromagnetically coupled (see Fig. \ref{fig:FIG6orbitals}b) and the out-of-plane \Fe\ and \Mn\ are antiferromagnetically coupled (see Fig. \ref{fig:FIG6orbitals}c). Because the lattice parameter $c_{pc}$ is just slightly bigger than  $a_{pc}$, ferroelectric displacements of \Fe\ and \Mn\ may have a significant effect on the energy levels of respective $d_{x^2-y^2}$ and $d_{z^2}$ orbitals. Especially in that case as illustrated in Fig. \ref{fig:FIG6orbitals}f,g,h where the $d_{x^2-y^2}$ and $d_{z^2}$ levels of \Mn\ are reversed, consequently the in-plane superexchange of \Fe\ and \Mn\ is switched from FM coupling (Fig. \ref{fig:FIG6orbitals}b) to AFM coupling (Fig. \ref{fig:FIG6orbitals}g), and meanwhile the out-of-plane superexchange of \Fe\ and \Mn\ is switched from AFM coupling (Fig. \ref{fig:FIG6orbitals}c) to FM coupling (Fig. \ref{fig:FIG6orbitals}h). Besides, whatever the \Fe-\Mn exchange interaction is, \Fe-\Fe and \Mn-\Mn exchange interactions favor AFM coupling because of the virtual hoping between the half-filled orbitals as given in Fig. \ref{fig:FIG6orbitals}d,e,f. Therefore, we can predict that there can exist an intermediate state with certain ferroelectric displacement, where the nearest neighbouring (NN) exchange interactions of \Fe-\Mn\ will become very small owning to the collective modulation of crystal field and ferroelectric distortions. In this situation, the next nearest neighbouring (NNN) exchange interactions of \Fe-\Fe\ and \Mn-\Mn\ become dominant and determine the magnetic state.\\

\begin{figure}[!ht]
\centering
\includegraphics[width=34pc,clip]{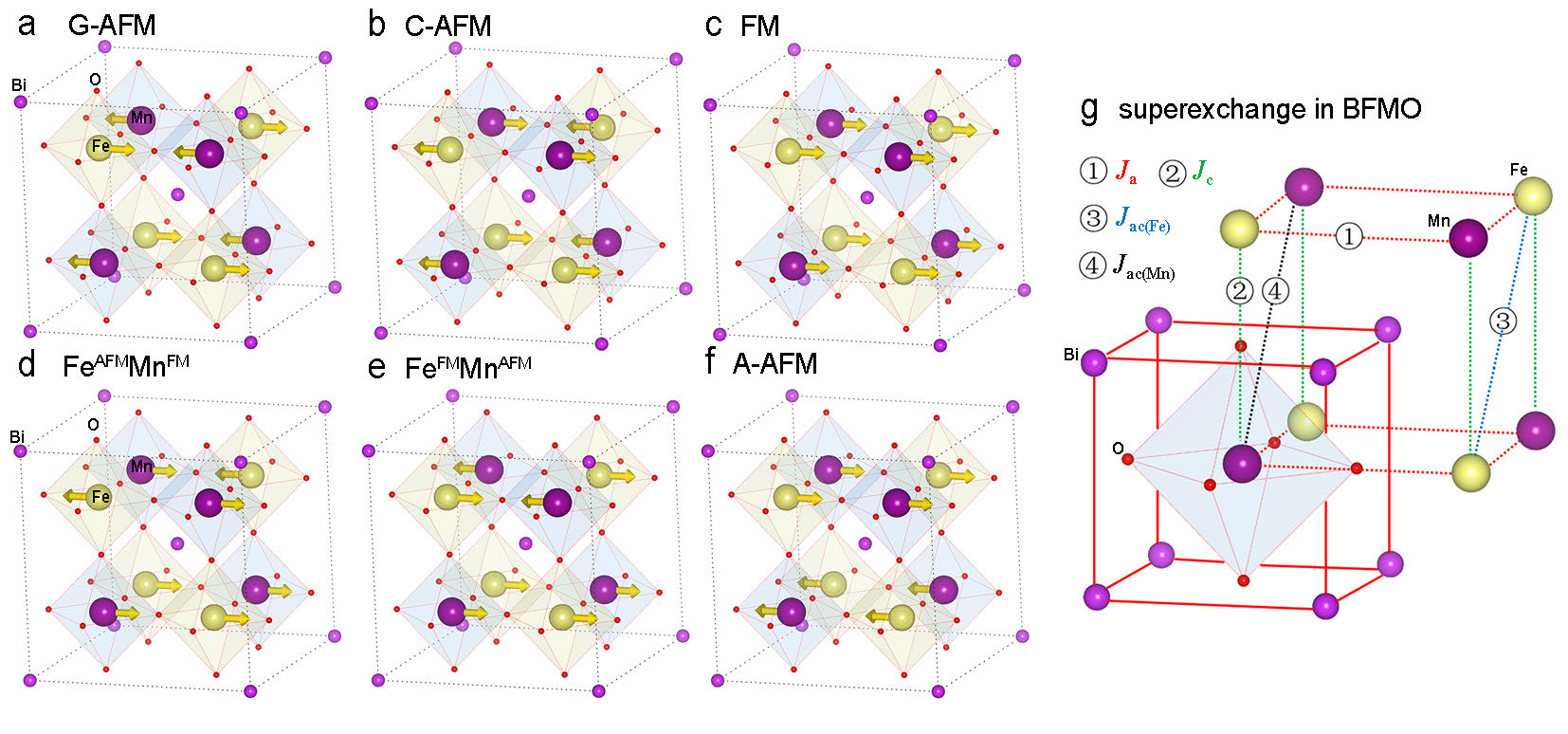}
\caption{\textbf{Schematic view of the magnetic structures and the superexchange considered in our models.} (a) The magnetic moments are ferromagnetically coupled within the (111) planes and antiferromagnetically coupled between adjacent planes, showing G-AFM magnetic order. (b) C-AFM refers to AFM coupling for (100) and (010) planes while the magnetic moments are ferromagnetically coupled within (110) planes. (c) In FM spin configuration, all of the magnetic moments are coupled ferromagnetically. (d) Fe$\rm ^{AFM}$Mn$\rm ^{FM}$ has AFM ordered Fe sublattice and FM ordered Mn sublattice. (e) Fe$\rm ^{FM}$Mn$\rm ^{AFM}$ has FM ordered Fe sublattice and AFM ordered Mn sublattice. (f) In A-AFM magnetic structure, the magnetic moments are ferromagnetically coupled within the (001) planes and antiferromagnetically coupled between adjacent planes. Arrows denote the magnetic moments of Fe and Mn. (g)The superexchange considered in our study. \textcircled{1} and \textcircled{2} denote the in-plane NN Fe-Mn exchange parameter $J\rm _a$ and the out-of-plane NN Fe-Mn exchange $J\rm _c$, respectively. \textcircled{3} and \textcircled{4} represent the NNN Fe-Fe exchange parameter $J\rm _{ac(Fe)}$ and the NNN Mn-Mn exchange parameter $J\rm _{ac(Mn)}$, respectively.} \label{fig:FIG7spinconfig}
\end{figure}

 Indeed, as proved in our first-principles calculations, we find a ferroelectric phase in which the superexchange of NNN \Fe-\Fe\ and \Mn-\Mn\ are much bigger than the NN \Fe-\Mn\ superexchange. Six magnetic structures (see Fig. \ref{fig:FIG7spinconfig}a-f) are adopted to perform total energy calculations and Heisenberg Hamiltonian is used to describe the exchange parameters, as shown in Fig. \ref{fig:FIG7spinconfig}g. The computational methods and model details can be found in \textbf{Methods section: \FP{} calculations}. The obtained $J\rm _a$, $J\rm _c$, $J\rm _{ac(Fe)}$ and $J\rm _{ac(Mn)}$ are -2.38 meV (FM), 1.18 meV (AFM), 10.50 meV (AFM) and 4.02 meV (AFM), respectively. One can notice that $J\rm _{ac(Fe)}$ is 4.4 times and 8.89 times as big as $J\rm _a$ and $J\rm _c$, respectively. These relatively bigger long-range AFM interactions give rise to the spin frustration in their respective magnetic sublattices and finally leads to the glassy state as we have observed in the experiment. Further quantitative analysis about how the ferroelectric distortion and exchange interactions are coupled will be discussed elsewhere\cite{FANG.unpublished}{}. \\

Using the parameters obtained in first-principles calculations, Monte Carlo simulations are used to predict the magnetic state at low temperature. The simulation methods can be found in \textbf{Methods section: Monte Carlo simulations}. The final magnetic state resulted from our simulations is represented by a three-dimensional model, as displayed in Fig. S6. One can see that it is a very frustrated system where spin directions are arranged randomly. In this system, random spins cancel each other, consequently the net magnetization is weak. In simulations, we also find that the magnetic structure is extremely sensitive to perturbations, e.g. external magnetic field, which is one of the most notable features of glassy state. These theoretical results agrees well with our experimental observations.\\

\begin{figure}[!ht]
\centering
\includegraphics[width=28pc,clip]{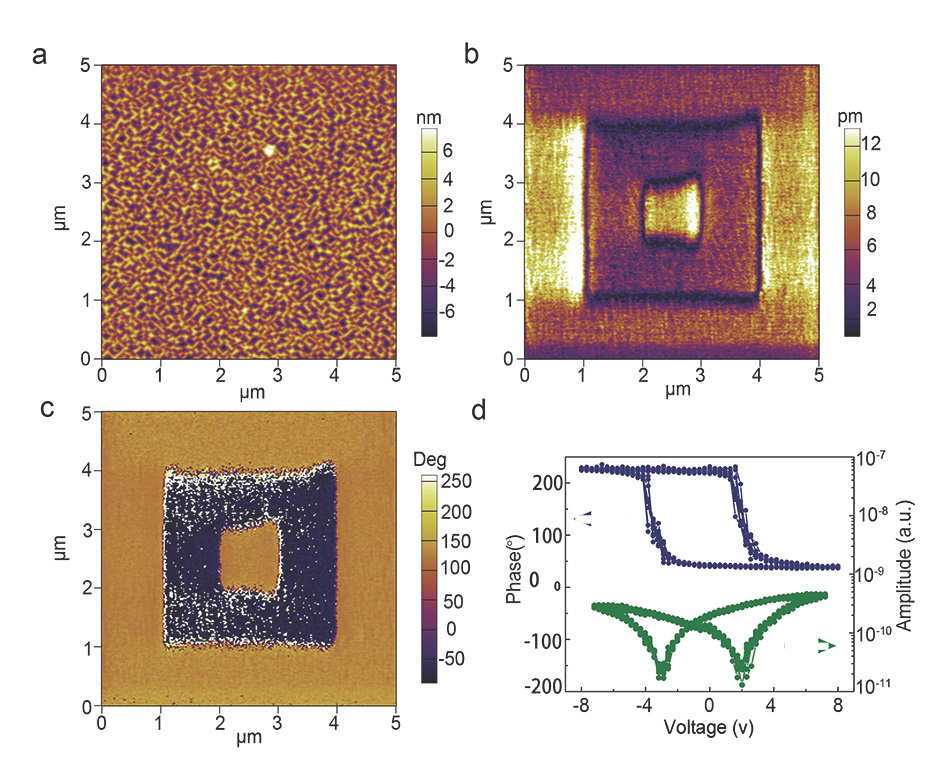}
\caption{\textbf{Local ferroelectric switching of BFMO film at room temperature.} (a) Surface morphology of BFMO film. (b) The out-of-plane PFM amplitude and phase (c) image recorded after writing using a conductive tip. (d) Local PFM hysteresis loops: phase signal (top); amplitude signal (bottom).} \label{fig:FIG8PFM}
\end{figure}

\textbf{Ferroelectricity characterization and analysis.} To identify if the RT ferroelectricity remains in BFMO system, we have further investigated the RT ferroelectricity of BFMO film on conductive SrRuO$_3$-coated STO (001) substrate and the external electric field can be applied on BFMO film easily. Piezoelectric Force Microscopy (PFM) was used to probe the local ferroelectric properties of this film. BFMO film on SrRuO$_3$/STO has similar surface morphology with that of BFMO film directly deposited on STO, as shown in Fig. \ref{fig:FIG8PFM}a. Fig. \ref{fig:FIG8PFM}b,c present the out-of-plane magnitude and phase image of ferroelectric domain after box-in-box switching with a tip bias of $\pm$ 9 V on the surface of BFMO film. A clear domain pattern after switching indicates the presence of RT ferroelectricity in BFMO film. For example, the yellow and dark colors in phase image represent the polarization state points downwards and upwards, respectively. The approximate 180$^\circ$ phase contrast reveals the two domains have entirely opposite polarization directions. Local PFM hysteresis loop in Fig. \ref{fig:FIG8PFM}d further confirms the existence of switching polarization, and good RT ferroelectric character in BFMO film. The local coercive voltages are about -3 V and 2 V, which are also indicated by the minimum value of the amplitude loop.\\

\section*{Discussion}
In summary, we have successfully fabricated high-quality BFMO films along [111] Fe-Mn ordering at RT temperature, which is unambiguously supported by our synchrotron XRD and HRTEM results. The exotic aspect of the present finding is the coexistence of spin glass state and ferroelectricity, which is suggested by SQUID and XMCD magnetic characterization, together with PFM analysis for ferroelectricity. Our \fp{} calculations proposes a spin frustration model to explain the observed glassy state and weak magnetism by considering the robust next nearest neighboring antiferromagnetic superexchange interactions, and further reveals a close correlation between ferroelectric distortions and exchange interactions which hints a notable magnetoelectric effect. Our study paves a way to attain complex magnetic structures in metastable complex oxides by epitaxial techniques. Studies on these complex oxides certainly have deeply enriched our understanding on the internal relationship between several order parameters, such as polarization and magnetization.
\section*{Methods}
\textbf{Detailed growth conditions.} A Bi$_{2.1}$FeMnO$_6$ ceramic target was synthesized by the conventional solid-state reaction method. 40 and 190 nm thick BFMO epitaxial films were deposited on STO (001) substrates from the Bi$_{2.1}$FeMnO$_6$ target by PLD, respectively. An excimer KrF ($\lambda$ = 248 nm) laser at a repetition rate of 8 Hz was used and the laser fluence on target was fixed to $\sim$ 2 J$\cdot$cm$^{-2}$. The BFMO films were prepared in 700 $^{\circ}$C and under 15 mTorr oxygen atmospheres.\\
\textbf{XRD measurements.} 2theta-Omega Scan, Rocking Curve Scan, Asymmetric (111) Scan, and RSM of BFMO epitaxial films were performed by high-resolution XRD (Bruker D8, Model: Discovery). A wavelength value $\lambda$ = 0.15406 nm for the CuK$\alpha$1 radiation and Ge (220) monochromator were employed in the XRD measurement.  The thickness of 40 nm BFMO film was determined by both XRR and the fringe-peak oscillation. RSM on (111) and (1/2,1/2,1/2) reflections experiments of BFMO films were identified using high resolution synchrotron X-ray diffractometry at the BL14B1-XRD beam line of Shanghai Synchrotron Radiation Facility.\\
\textbf{TEM image and SAED conditions.} HRTEM cross-section image of BFMO and Selected Area Electron Diffraction were performed by Field Emission Transmission Electron Microscopy (2100F, JEOL, Co., Tokyo, Japan).\\
\textbf{Atomic force microscopy and PFM measurements.} The surface morphology of BFMO films were observed by Atomic Force Microscopy (Veeco Dimension 3100). Local ferroelectric properties of BFMO films were probed by Piezoelectric Force Microscopy (PFM, Asylum Research Cypher).\\
\textbf{XPS measurement.} X-ray Photoelectron Spectroscopy (XPS, Shimadzu-Kratos, Model: AXIS Ultra$^{DLD}$) was used to identify the chemical valence state of all metallic elements in BFMO films. The binding energies were calibrated by using the carbon peak (C 1s = 284.8 eV).\\
\textbf{SQUID and PPMS measurements.} Magnetism of BFMO films were measured by Magnetic Properties Measurement System (MPMS3, Quantum Design) and Physical Properties Measurement Systems (PPMS-9, Quantum Design).\\
\textbf{XAS and XMCD measurements.} Synchrotron X-ray absorption based measurements were performed at Beamline 11A of National Synchrotron Radiation Research Center in Hsinchu, Taiwan. Circularly polarized X-rays were selected with fixed polarization. $\pm$ 1 Tesla external magnetic field were applied during the XMCD measurements. All measurements were done at room temperature.\\
\textbf{\FP{} calculations.} \emph{Computational methods}: Our calculations were performed by both full-electron calculations and projected augmented wave (PAW) method. All-electron calculations use the full-potential linearised augmented-plane wave method, as implemented in the WIEN2k code\cite{Blaha.WIEN2k.2001}{}, and PAW calculations use the plane wave basis set in Vienna $ab$-initio simulation package (VASP)\cite{Kresse.PRB.1996}{}. Generalized gradient approximation (GGA) and GGA plus Hubbard $U$ (GGA+$U$) were treated for the exchange-correlation potential. We employed effective Hubbard $U$ of 4 eV and 5.2 eV\cite{Rabe.PRL.2010,Rabe.PRB.2010} for Fe and Mn 3$d$ orbitals, respectively. In VASP\cite{Kresse.PRB.1996}{}, the energy cutoff for the plane-wave basis was set to 520 eV in relaxations and 800 eV in total energy calculations, respectively. $8 \times 8 \times 6$ k-point meshes were adopted to sample the first Brillouin-zone for BFMO. In WIEN2k\cite{Blaha.WIEN2k.2001}{}, the muffin-tin sphere radii are chosen to be 2.44, 1.9, 1.86 and 1.6 Bohr for Bi, Fe, Mn and O atoms, respectively. The cutoff energy of 16 Ry is used for plane wave expansion of interstitial wave functions, and $6 \times 6 \times 4$ k mesh for integration over the Brillouin zone. \emph{Computational models}: We built a $\sqrt{2}$ $\times$ $\sqrt{2}$  $\times$ 2 superlattice according to the experimental lattice constants. We simulated a set of structures with different ferroelectric distortions ($\lambda$). The results indicate that the exchange constants are closely correlated to the distortions. What is more, the spin frustration does not appear in the fully relaxed ferroelectric phase (i.e., $\lambda$ = 1). Only in certain phase with specific ferroelectric displacements ($\lambda$ $\approx$ 0.4 in our case) can the NNN AFM be enhanced to cause the magnetic frustration. We argue that such a phenomenon could be induced by the closely coupling of exchange interactions and ferroelectric polarization, hinting a possible magnetoelectric effect. \\
\textbf{Monte Carlo simulations.} Monte Carlo simulations based on the model Heisenberg Hamiltonian are used to simulate the glassy state. The same method has been applied successfully to predict the ground state of magnetic materials in our previous study\cite{Mont.PRL.2005}{}. In our simulations, we employ a 16a $\times$ 16a $\times$ 16c (a = 3.905, c = 4.015 \AA) face-centered cubic cell with periodic boundary conditions. For the sake of simplicity in simulations, Fe and Mn atoms are treated equally.


\section*{Acknowledgments}
This work was sponsored by the 973 Programs (No. 2014CB921104, 2013CB922301), the NSF of China (No. 61125403, 51572085, 61574057). Shanghai Synchrotron Radiation Facility (BL14B bean line) is greatly acknowledged for providing the beam time and technical assistance. All computations are conducted in computing center at ECNU and Chinese Tianhe-1A system at the National Supercomputer Center. Y.W.F. acknowledges Hung T. Diep, Tapan Chatterji and Xian-Gang Wan for their expert comments on spin frustration and exchange interactions. Sincere thanks also go to Yong-Ping Du, Hong Jian Zhao and Wei-Jia Fan for their respective assistance in calculations.
\section*{Author contributions}
$^{\dagger}$The authors L.S. and Y.W.F. contributed equally to this work. C.G.D. and L.S. conceived this study. L.S. and J.H. fabricated the samples. Y.W.F. and C.G.D. contributed to the orbital model analysis, first-principles calculations and Monte Carlo simulations. L.S. performed the RSM and PFM measurement. L.S. and Y.Z. analyzed the results of SQUID and PPMS measurement. R.H. and R.Q. supported TEM measurement. Q.H. and Y.H.C. provided XMCD data. All authors were involved in data interpretation and discussion. L.S., Y.W.F., R.H. and C.G.D. co-wrote this paper.

\textbf{Supplementary Figures}
\begin{figure}[!ht]
\centering
\includegraphics[width=34pc,clip]{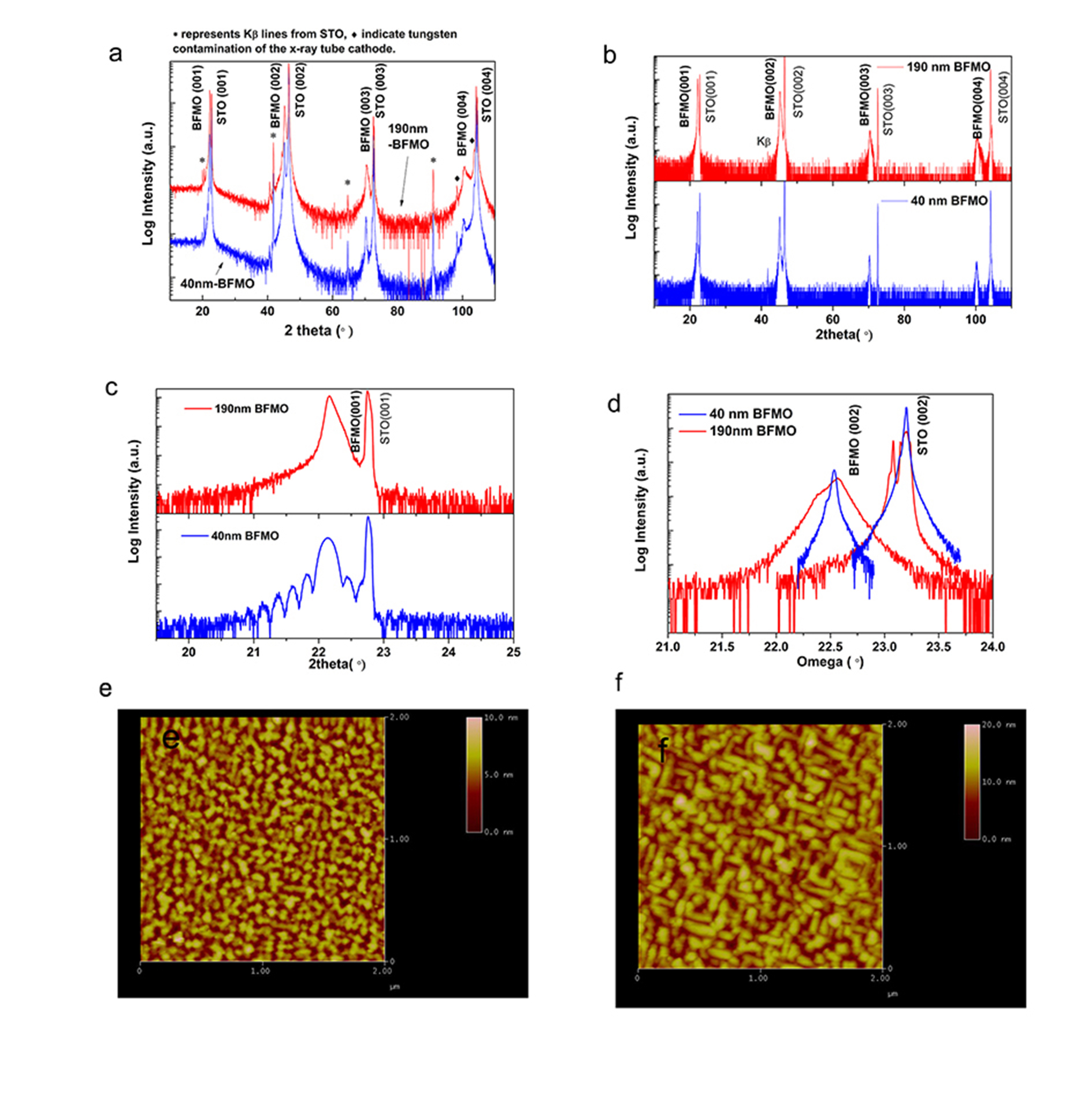}
\begin{flushleft}
\textbf{Supplementary Figure S1. Structural and topography of BFMO films.} (a)2theta-Omega XRD scans for BFMO films epitaxially grown on STO(001) substrates with Cu-K$_{\rm \alpha}$ radiation X-ray rather than monochromator-filtered Cu-K$_{\rm \alpha 1}$  in order to detect if impurity phase exists in BFMO films. High-Resolution XRD 2theta-Omega scan of BFMO films grown on STO(001) substrates, (b) wide range: 10$^\circ$ $\sim$ 110$^\circ$, (c) in the vicinity of (001) reflections. (d) rocking curve scan of BFMO films and STO substrates around (002) reflections. A topography of (e) 40 nm BFMO film and (f) 190 nm BFMO film obtained by Atomic Force Microscopy.
\end{flushleft} \label{fig:FIGS1}
\end{figure}

\begin{figure}[!ht]
\centering
\includegraphics[width=34pc,clip]{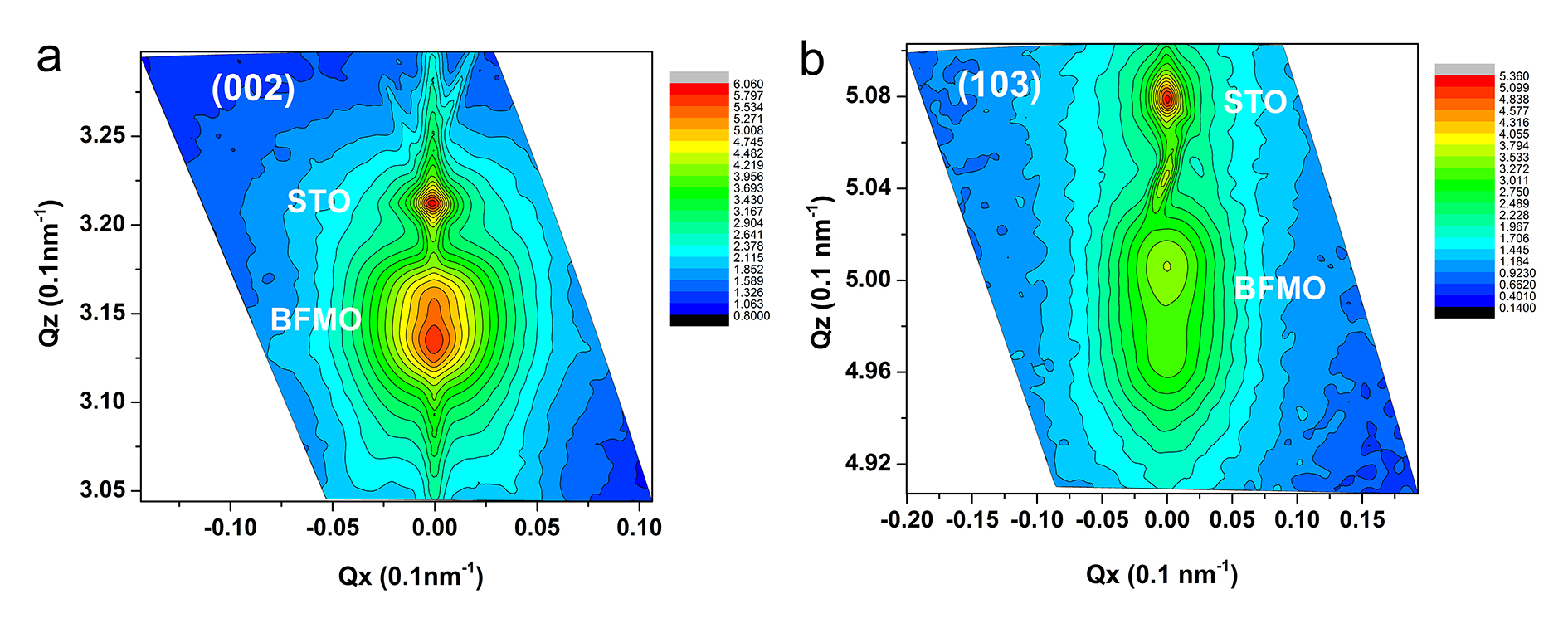}
\begin{flushleft}
\textbf{Supplementary Figure S2. eciprocal Space Mapping of 190 nm BFMO film using synchrotron X-ray diffraction.} (a) Reciprocal Space Mapping of (002) reflection of 190 nm-thick BFMO film and STO substrate. (b) Reciprocal space mapping (103) reflection of 190 nm-thick BFMO film and STO substrate. \end{flushleft} \label{fig:FIGS2}
\end{figure}

\begin{figure}[!ht]
\centering
\includegraphics[width=34pc,clip]{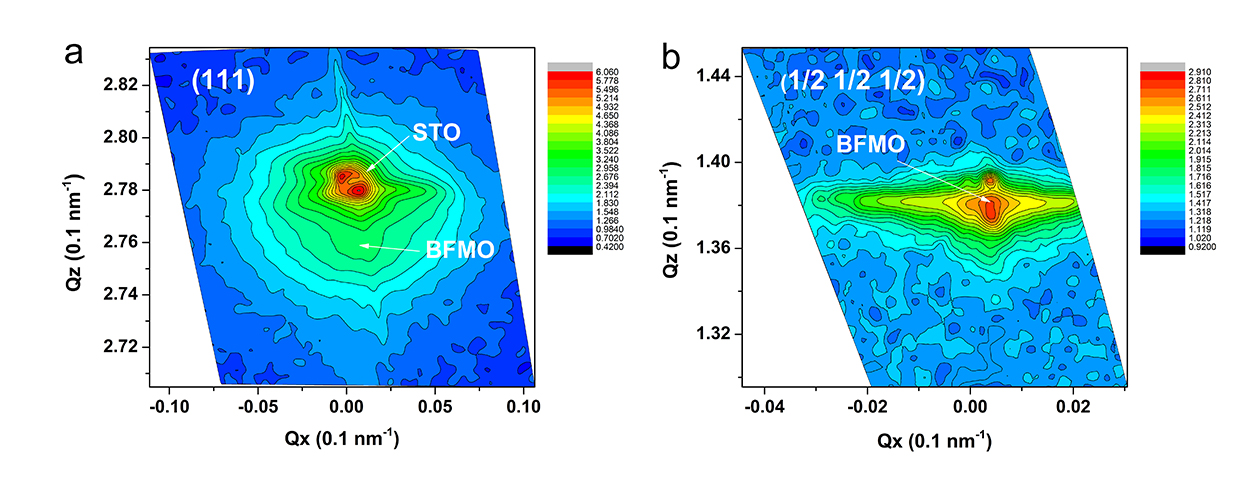}
\begin{flushleft}
\textbf{Supplementary Figure S3. Direct identification of Fe and Mn ordering along [111] direction using synchrotron X-ray diffraction.} (a) Reciprocal space mapping of (111) reflection of 40 nm-thick BFMO film and STO substrate. (b) Reciprocal space mapping of superlattice (1/2, 1/2, 1/2) reflection of 40 nm-thick BFMO film.
\end{flushleft}  \label{fig:FIGS3}
\end{figure}

\begin{figure}[!ht]
\centering
\includegraphics[width=34pc,clip]{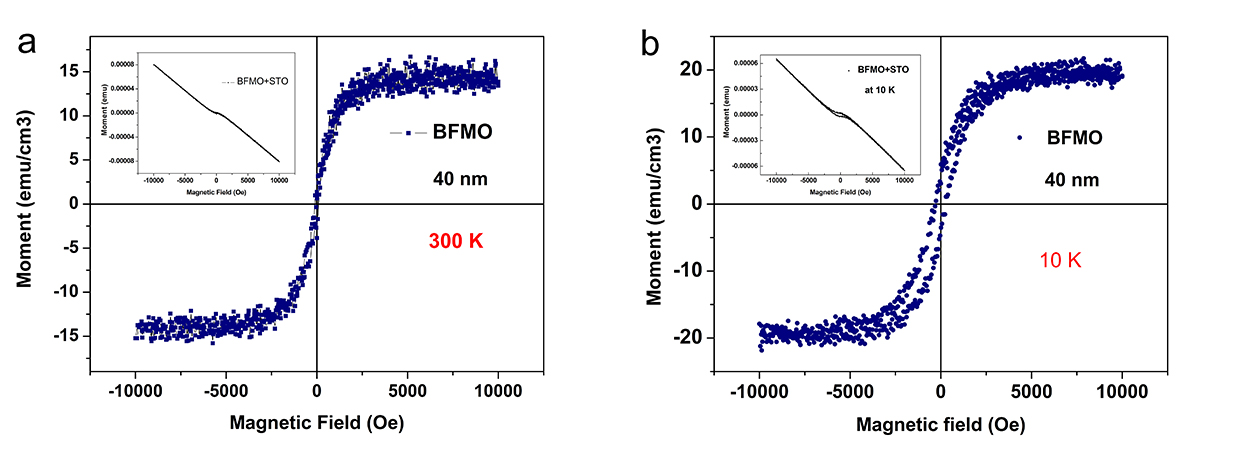}
\begin{flushleft}
\textbf{Supplementary Figure S4. Magnetic characterization of BFMO film.} In-plane magnetic hysteresis $M$-$H$ curves of 40 nm-thick BFMO film epitaxially grown on STO (001) substrate after the subtraction of the diamagnetic signal of STO substrate, (a) measured at 300 K, (b) at 10 K. These M-H curves were measured by Physical Properties Measurement Systems (PPMS).
\end{flushleft}  \label{fig:FIGS4}
\end{figure}

\begin{figure}[!ht]
\centering
\includegraphics[width=34pc,clip]{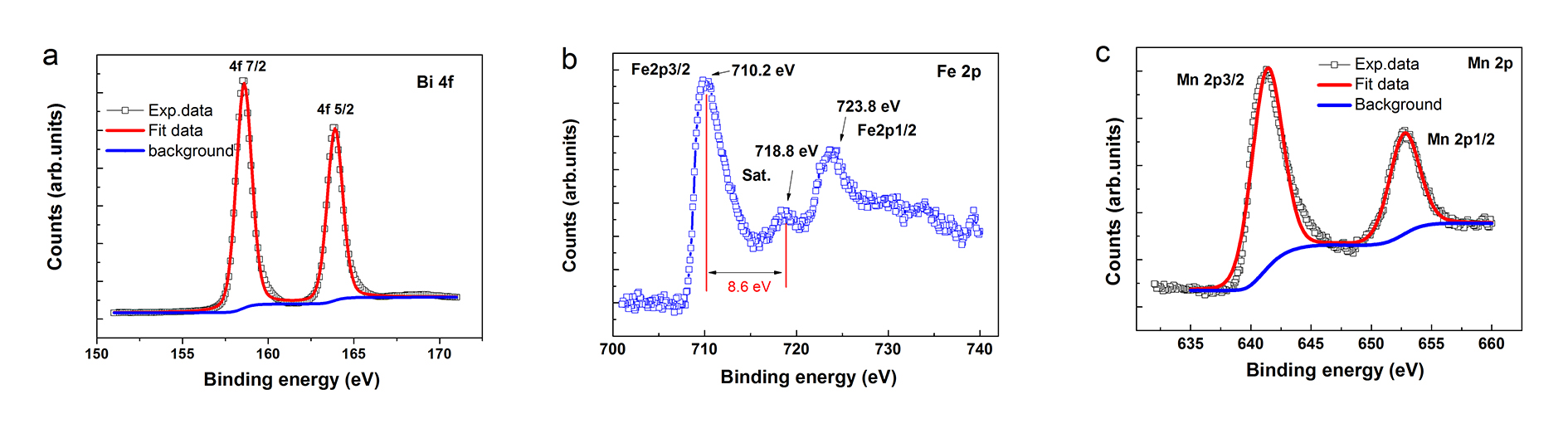}
\begin{flushleft}
\textbf{Supplementary Figure S5. Characterization of element valence state in BFMO. } XPS spectra of Bi 4$f$ (a) , Fe 2$p$ (b) and Mn 2$p$ (c) core levels for 190 nm-thick BFMO film deposited on STO (001) substrate.
\end{flushleft}  \label{fig:FIGS5}
\end{figure}

\begin{figure}[!ht]
\centering
\includegraphics[width=34pc,clip]{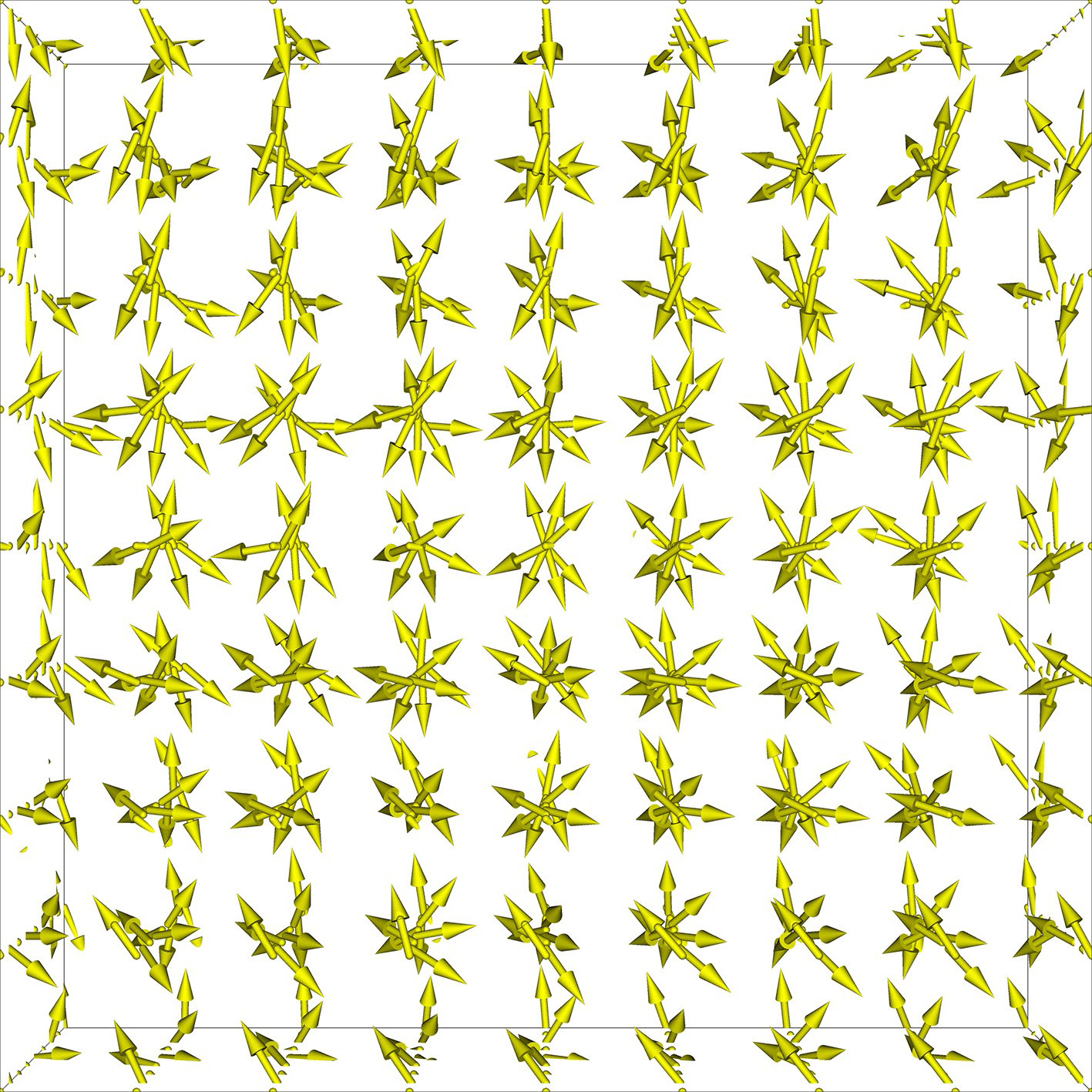}
\begin{flushleft}
\textbf{Supplementary Figure S6. The spin glass state resulted in Monte Carlo simulations. } Here, we plot the range $8\times8\times8$ and set the view point at (0, 0, 4). The arrows indicate the spin directions of Fe/Mn ions.
\end{flushleft}  \label{fig:FIGS6}
\end{figure}


\begin{thebibliography}{10}
\expandafter\ifx\csname url\endcsname\relax
  \def\url#1{\texttt{#1}}\fi
\expandafter\ifx\csname urlprefix\endcsname\relax\def\urlprefix{URL }\fi
\providecommand{\bibinfo}[2]{#2}
\providecommand{\eprint}[2][]{\url{#2}}

\bibitem{Bibes.Nat.Mat.2008}
\bibinfo{author}{Bibes, M.} \& \bibinfo{author}{Barthelemy, A.}
\newblock \bibinfo{title}{Multiferroics: Towards a magnetoelectric memory}.
\newblock \emph{\bibinfo{journal}{Nature Mater.}}
  \textbf{\bibinfo{volume}{7}}, \bibinfo{pages}{425--426}
  (\bibinfo{year}{2008}).

\bibitem{Cheong.NatMat.2007}
\bibinfo{author}{Cheong, S.-W.} \& \bibinfo{author}{Mostovoy, M.}
\newblock \bibinfo{title}{Multiferroics: a magnetic twist for
  ferroelectricity}.
\newblock \emph{\bibinfo{journal}{Nature Mater.}}
  \textbf{\bibinfo{volume}{6}}, \bibinfo{pages}{13--20} (\bibinfo{year}{2007}).

\bibitem{Ramesh.NatMat.2007}
\bibinfo{author}{Ramesh, R.} \& \bibinfo{author}{Spaldin, N.~A.}
\newblock \bibinfo{title}{Multiferroics: progress and prospects in thin films}.
\newblock \emph{\bibinfo{journal}{Nature Mater.}}
  \textbf{\bibinfo{volume}{6}}, \bibinfo{pages}{21--29} (\bibinfo{year}{2007}).

\bibitem{Spaldin.JPCB.2000}
\bibinfo{author}{Hill, N.~A.}
\newblock \bibinfo{title}{Why are there so few magnetic ferroelectrics?}
\newblock \emph{\bibinfo{journal}{J. Phys. Chem. B}}
  \textbf{\bibinfo{volume}{104}}, \bibinfo{pages}{6694--6709}
  (\bibinfo{year}{2000}).

\bibitem{Smolenskii.ferro.1982}
\bibinfo{author}{Smolenskii, G.~A.} \& \bibinfo{author}{Chupis, I.~E.}
\newblock \bibinfo{title}{Ferroelectromagnets}.
\newblock \emph{\bibinfo{journal}{Sov. Phys. Usp.}}
  \textbf{\bibinfo{volume}{25}}, \bibinfo{pages}{475} (\bibinfo{year}{1982}).

\bibitem{Darrell.APL.2004}
\bibinfo{author}{Moreira~dos Santos, A.~F.} \emph{et~al.}
\newblock \bibinfo{title}{Epitaxial growth and properties of metastable
  {BiMnO$_{3}$} thin films}.
\newblock \emph{\bibinfo{journal}{Appl. Phys. Lett.}}
  \textbf{\bibinfo{volume}{84}}, \bibinfo{pages}{91--93}
  (\bibinfo{year}{2004}).

\bibitem{Faqir.JSSC.1999}
\bibinfo{author}{Faqir, H.} \emph{et~al.}
\newblock \bibinfo{title}{High-temperature xrd and dta studies of {BiMnO$_{3}$}
  perovskite}.
\newblock \emph{\bibinfo{journal}{J. Solid. State. Chem.}}
  \textbf{\bibinfo{volume}{142}}, \bibinfo{pages}{113--119}
  (\bibinfo{year}{1999}).

\bibitem{Lee.APL.2010}
\bibinfo{author}{Lee, J.~H.} \emph{et~al.}
\newblock \bibinfo{title}{Adsorption-controlled growth of {BiMnO$_{3}$} films
  by molecular-beam epitaxy}.
\newblock \emph{\bibinfo{journal}{Appl. Phys. Lett.}}
  \textbf{\bibinfo{volume}{96}}, \bibinfo{pages}{262905}
  (\bibinfo{year}{2010}).

\bibitem{Luca.APL.2013}
\bibinfo{author}{De~Luca, G.~M.} \emph{et~al.}
\newblock \bibinfo{title}{Ferromagnetism and ferroelectricity in epitaxial
  {BiMnO$_{3}$} ultra-thin films}.
\newblock \emph{\bibinfo{journal}{Appl. Phys. Lett.}}
  \textbf{\bibinfo{volume}{103}}, \bibinfo{pages}{062902}
  (\bibinfo{year}{2013}).

\bibitem{BiLei.PRB.2008}
\bibinfo{author}{Bi, L.} \emph{et~al.}
\newblock \bibinfo{title}{Structural, magnetic, and optical properties of
  {BiFeO$_{3}$} and {Bi$_{2}$FeMnO$_{6}$} epitaxial thin films: An experimental
  and first-principles study}.
\newblock \emph{\bibinfo{journal}{Phys. Rev. B}}
  \textbf{\bibinfo{volume}{78}}, \bibinfo{pages}{104106}
  (\bibinfo{year}{2008}).

\bibitem{Zhao.APL.2009}
\bibinfo{author}{Zhao, H.~Y.}, \bibinfo{author}{Kimura, H.},
  \bibinfo{author}{Cheng, Z.~X.}, \bibinfo{author}{Wang, X.~L.} \&
  \bibinfo{author}{Nishida, T.}
\newblock \bibinfo{title}{Room temperature multiferroic properties of
  {Nd:BiFeO$_{3}$/Bi$_{2}$FeMnO$_{6}$} bilayered films}.
\newblock \emph{\bibinfo{journal}{Appl. Phys. Lett.}}
  \textbf{\bibinfo{volume}{95}}, \bibinfo{pages}{232904}
  (\bibinfo{year}{2009}).

\bibitem{DuYi.APL.2010}
\bibinfo{author}{Du, Y.} \emph{et~al.}
\newblock \bibinfo{title}{Magnetic properties of {Bi$_{2}$FeMnO$_{6}$}: A
  multiferroic material with double-perovskite structure}.
\newblock \emph{\bibinfo{journal}{Appl. Phys. Lett.}}
  \textbf{\bibinfo{volume}{97}}, \bibinfo{pages}{122502}
  (\bibinfo{year}{2010}).

\bibitem{Choi.AFM.2014}
\bibinfo{author}{Choi, E.-M.} \emph{et~al.}
\newblock \bibinfo{title}{Room temperature ferrimagnetism and ferroelectricity
  in strained, thin films of {BiFe$_{0.5}$Mn$_{0.5}$O$_{3}$}}.
\newblock \emph{\bibinfo{journal}{Adv. Funct. Mater.}}
  \textbf{\bibinfo{volume}{24}}, \bibinfo{pages}{7478--7487}
  (\bibinfo{year}{2014}).

\bibitem{Rabe.PRB.2010}
\bibinfo{author}{P$\acute{a}$lov$\acute{a}$, L.}, \bibinfo{author}{Chandra, P.}
  \& \bibinfo{author}{Rabe, K.~M.}
\newblock \bibinfo{title}{Multiferroic {BiFeO$_{3}$}-{BiMnO$_{3}$} nanoscale
  checkerboard from first principles}.
\newblock \emph{\bibinfo{journal}{Phys. Rev. B}}
  \textbf{\bibinfo{volume}{82}}, \bibinfo{pages}{075432}
  (\bibinfo{year}{2010}).

\bibitem{Rabe.PRL.2010}
\bibinfo{author}{P$\acute{a}$lov$\acute{a}$, L.}, \bibinfo{author}{Chandra, P.}
  \& \bibinfo{author}{Rabe, K.~M.}
\newblock \bibinfo{title}{Magnetostructural effect in the multiferroic
  {BiFeO$_{3}$}--{BiMnO$_{3}$} checkerboard from first principles}.
\newblock \emph{\bibinfo{journal}{Phys. Rev. Lett.}}
  \textbf{\bibinfo{volume}{104}}, \bibinfo{pages}{037202}
  (\bibinfo{year}{2010}).

\bibitem{Kundu.APL.2008}
\bibinfo{author}{Kundu, A.~K.} \emph{et~al.}
\newblock \bibinfo{title}{A multiferroic ceramic with perovskite
  structure:{(La$_{0.5}$Bi$_{0.5}$)(Mn$_{0.5}$Fe$_{0.5}$)O$_{3.09}$}}.
\newblock \emph{\bibinfo{journal}{Appl. Phys. Lett.}}
  \textbf{\bibinfo{volume}{93}}, \bibinfo{pages}{052906}
  (\bibinfo{year}{2008}).

\bibitem{Mandal.PRB.2010}
\bibinfo{author}{Mandal, P.} \emph{et~al.}
\newblock \bibinfo{title}{Temperature-induced magnetization reversal in
  {BiFe$_{0.5}$Mn$_{0.5}$O$_{3}$} synthesized at high pressure}.
\newblock \emph{\bibinfo{journal}{Phys. Rev. B}}
  \textbf{\bibinfo{volume}{82}}, \bibinfo{pages}{100416}
  (\bibinfo{year}{2010}).

\bibitem{Delmonte.PRB.2013}
\bibinfo{author}{Delmonte, D.} \emph{et~al.}
\newblock \bibinfo{title}{Thermally activated magnetization reversal in bulk
  {BiFe$_{0.5}$Mn$_{0.5}$O$_{3}$}}.
\newblock \emph{\bibinfo{journal}{Phys. Rev. B}}
  \textbf{\bibinfo{volume}{88}}, \bibinfo{pages}{014431}
  (\bibinfo{year}{2013}).

\bibitem{Zhao.JAP.2010}
\bibinfo{author}{Zhao, H.~Y.} \emph{et~al.}
\newblock \bibinfo{title}{Magnetic properties of la doped {Bi$_{2}$FeMnO$_{6}$}
  ceramic and film}.
\newblock \emph{\bibinfo{journal}{J. Appl. Phys.}}
  \textbf{\bibinfo{volume}{108}}, \bibinfo{pages}{093903}
  (\bibinfo{year}{2010}).

\bibitem{Zhao.2010}
\bibinfo{author}{Zhao, H.} \emph{et~al.}
\newblock \bibinfo{title}{Magnetic characterization of bi2femno6 film grown on
  (100) {SrTiO$_{3}$} substrate}.
\newblock \emph{\bibinfo{journal}{phys. status solidi RRL}} \textbf{\bibinfo{volume}{4}}, \bibinfo{pages}{314--316}
  (\bibinfo{year}{2010}).

\bibitem{Cortie.APL.2012}
\bibinfo{author}{Cortie, D.} \emph{et~al.}
\newblock \bibinfo{title}{The magnetic structure of an epitaxial
  {BiMn$_{0.5}$Fe$_{0.5}$O$_{3}$} thin film on {SrTiO$_{3}$} (001) studied with
  neutron diffraction}.
\newblock \emph{\bibinfo{journal}{Appl. Phys. Lett.}}
  \textbf{\bibinfo{volume}{101}}, \bibinfo{pages}{172404}
  (\bibinfo{year}{2012}).

\bibitem{Liu.JAP.2013}
\bibinfo{author}{Liu, P.} \emph{et~al.}
\newblock \bibinfo{title}{Anisotropy of crystal growth mechanisms,
  dielectricity, and magnetism of multiferroic {Bi$_{2}$FeMnO$_{6}$} thin films}.
\newblock \emph{\bibinfo{journal}{J. Appl. Phys.}}
  \textbf{\bibinfo{volume}{113}}, \bibinfo{pages}{17D904}
  (\bibinfo{year}{2013}).

\bibitem{Choi.APL.2011}
\bibinfo{author}{Choi, E.-M.} \emph{et~al.}
\newblock \bibinfo{title}{Strong room temperature magnetism in highly resistive
  strained thin films of {BiFe$_{0.5}$Mn$_{0.5}$O$_{3}$}}.
\newblock \emph{\bibinfo{journal}{Appl. Phys. Lett.}}
  \textbf{\bibinfo{volume}{98}}, \bibinfo{pages}{012509}
  (\bibinfo{year}{2011}).

\bibitem{Miao.APL.2011}
\bibinfo{author}{Miao, J.}, \bibinfo{author}{Zhang, X.}, \bibinfo{author}{Zhan,
  Q.}, \bibinfo{author}{Jiang, Y.} \& \bibinfo{author}{Chew, K.-H.}
\newblock \bibinfo{title}{Bi-relaxation behaviors in epitaxial multiferroic
  double-perovskite {BiFe$_{0.5}$Mn$_{0.5}$O$_{3}$/CaRuO$_{3}$}
  heterostructures}.
\newblock \emph{\bibinfo{journal}{Appl. Phys. Lett.}}
  \textbf{\bibinfo{volume}{99}}, \bibinfo{pages}{062905}
  (\bibinfo{year}{2011}).

\bibitem{ChenAiping.AM.2013}
\bibinfo{author}{Chen, A.} \emph{et~al.}
\newblock \bibinfo{title}{A new class of room-temperature multiferroic thin
  films with bismuth-based supercell structure}.
\newblock \emph{\bibinfo{journal}{Adv. Mater.}}
  \textbf{\bibinfo{volume}{25}}, \bibinfo{pages}{1028--1032}
  (\bibinfo{year}{2013}).

\bibitem{Nechache.JAP.2009}
\bibinfo{author}{Nechache, R.} \emph{et~al.}
\newblock \bibinfo{title}{Epitaxial thin films of the multiferroic double
  perovskite {Bi$_{2}$FeCrO$_{6}$} grown on (100)-oriented {SrTiO$_{3}$}
  substrates: Growth, characterization, and optimization}.
\newblock \emph{\bibinfo{journal}{J. Appl. Phys.}}
  \textbf{\bibinfo{volume}{105}}, \bibinfo{pages}{061621}
  (\bibinfo{year}{2009}).

\bibitem{Nechache.JPCM.2012}
\bibinfo{author}{Nechache, R.}, \bibinfo{author}{Harnagea, C.} \&
  \bibinfo{author}{Pignolet, A.}
\newblock \bibinfo{title}{Multiferroic properties-structure relationships in
  epitaxial {Bi$_{2}$FeCrO$_{6}$} thin films: recent developments}.
\newblock \emph{\bibinfo{journal}{J. Phys. Condens. Matter}}
  \textbf{\bibinfo{volume}{24}}, \bibinfo{pages}{096001}
  (\bibinfo{year}{2012}).

\bibitem{Nechache.APL.2006}
\bibinfo{author}{Nechache, R.} \emph{et~al.}
\newblock \bibinfo{title}{Growth, structure, and properties of epitaxial thin
  films of first-principles predicted multiferroic {Bi$_{2}$FeCrO$_{6}$}}.
\newblock \emph{\bibinfo{journal}{Appl. Phys. Lett.}}
  \textbf{\bibinfo{volume}{89}}, \bibinfo{pages}{102902}
  (\bibinfo{year}{2006}).

\bibitem{Nechache.Nat.Phot.2015}
\bibinfo{author}{R, N.} \emph{et~al.}
\newblock \bibinfo{title}{Bandgap tuning of multiferroic oxide solar cells}.
\newblock \emph{\bibinfo{journal}{Nature Photon.}}
  \textbf{\bibinfo{volume}{9}}, \bibinfo{pages}{61--67} (\bibinfo{year}{2015}).

\bibitem{Sakai.APL.2007}
\bibinfo{author}{Sakai, M.} \emph{et~al.}
\newblock \bibinfo{title}{Multiferroic thin film of {Bi$_{2}$NiMnO$_{6}$} with
  ordered double-perovskite structure}.
\newblock \emph{\bibinfo{journal}{Appl. Phys. Lett.}}
  \textbf{\bibinfo{volume}{90}}, \bibinfo{pages}{072903}
  (\bibinfo{year}{2007}).

\bibitem{Ueda.APL.2001}
\bibinfo{author}{Ueda, K.}, \bibinfo{author}{Muraoka, Y.},
  \bibinfo{author}{Tabata, H.} \& \bibinfo{author}{Kawai, T.}
\newblock \bibinfo{title}{Atomic ordering in the
  {LaFe$_{0.5}$Mn$_{0.5}$O$_{3}$} solid solution film}.
\newblock \emph{\bibinfo{journal}{Appl. Phys. Lett.}}
  \textbf{\bibinfo{volume}{78}}, \bibinfo{pages}{512--514}
  (\bibinfo{year}{2001}).

\bibitem{Ueda.PRB.1999}
\bibinfo{author}{Ueda, K.}, \bibinfo{author}{Tabata, H.} \&
  \bibinfo{author}{Kawai, T.}
\newblock \bibinfo{title}{Atomic arrangement and magnetic properties of
  {LaFeO$_{3}$--LaMnO$_{3}$} artificial superlattices}.
\newblock \emph{\bibinfo{journal}{Phys. Rev. B}}
  \textbf{\bibinfo{volume}{60}}, \bibinfo{pages}{R12561}
  (\bibinfo{year}{1999}).

\bibitem{Xu.APL.2010}
\bibinfo{author}{Xu, X.~S.} \emph{et~al.}
\newblock \bibinfo{title}{Tunable band gap in {Bi(Fe$_{1-x}$Mn$_{x}$)O$_{3}$}
  films}.
\newblock \emph{\bibinfo{journal}{Appl. Phys. Lett.}}
  \textbf{\bibinfo{volume}{96}}, \bibinfo{pages}{192901}
  (\bibinfo{year}{2010}).

\bibitem{Ederer.PRB.2005}
\bibinfo{author}{Ederer, C.} \& \bibinfo{author}{Spaldin, N.~A.}
\newblock \bibinfo{title}{Weak ferromagnetism and magnetoelectric coupling in
  bismuth ferrite}.
\newblock \emph{\bibinfo{journal}{Phys. Rev. B}}
  \textbf{\bibinfo{volume}{71}}, \bibinfo{pages}{060401}
  (\bibinfo{year}{2005}).

\bibitem{Lebeugle.PRL.2008}
\bibinfo{author}{Lebeugle, D.} \emph{et~al.}
\newblock \bibinfo{title}{Electric-field-induced spin flop in {BiFeO$_{3}$}
  single crystals at room temperature}.
\newblock \emph{\bibinfo{journal}{Phys. Rev. Lett.}}
  \textbf{\bibinfo{volume}{100}}, \bibinfo{pages}{227602}
  (\bibinfo{year}{2008}).

\bibitem{Albrecht.PRB.2010}
\bibinfo{author}{Albrecht, D.} \emph{et~al.}
\newblock \bibinfo{title}{Ferromagnetism in multiferroic {BiFeO$_{3}$} films: A
  first-principles-based study}.
\newblock \emph{\bibinfo{journal}{Phys. Rev. B}}
  \textbf{\bibinfo{volume}{81}}, \bibinfo{pages}{140401(R)}
  (\bibinfo{year}{2010}).

\bibitem{Scott.AM.2009}
\bibinfo{author}{Catalan, G.} \& \bibinfo{author}{Scott, J.~F.}
\newblock \bibinfo{title}{Physics and applications of bismuth ferrite}.
\newblock \emph{\bibinfo{journal}{Adv. Mater.}}
  \textbf{\bibinfo{volume}{21}}, \bibinfo{pages}{2463--2485}
  (\bibinfo{year}{2009}).

\bibitem{Giri.JAP.2006}
\bibinfo{author}{De, K.}, \bibinfo{author}{Thakur, M.}, \bibinfo{author}{Manna,
  A.} \& \bibinfo{author}{Giri, S.}
\newblock \bibinfo{title}{Unusual glassy states in
  {LaMn$_{0.5}$Fe$_{0.5}$O$_{3}$}: Evidence of two distinct dynamical freezing
  processes}.
\newblock \emph{\bibinfo{journal}{J. Appl. Phys.}}
  \textbf{\bibinfo{volume}{99}}, \bibinfo{pages}{013908}
  (\bibinfo{year}{2006}).

\bibitem{Singh.PRB.2008}
\bibinfo{author}{Singh, M.~K.}, \bibinfo{author}{Prellier, W.},
  \bibinfo{author}{Singh, M.~P.}, \bibinfo{author}{Katiyar, R.~S.} \&
  \bibinfo{author}{Scott, J.~F.}
\newblock \bibinfo{title}{Spin-glass transition in single-crystal
  {BiFeO$_{3}$}}.
\newblock \emph{\bibinfo{journal}{Phys. Rev. B}}
  \textbf{\bibinfo{volume}{77}}, \bibinfo{pages}{144403}
  (\bibinfo{year}{2008}).

\bibitem{Xu.Sci.rep.2015}
\bibinfo{author}{Xu, Q.} \emph{et~al.}
\newblock \bibinfo{title}{Magnetic interactions in
  {BiFe$_{0.5}$Mn$_{0.5}$O$_{3}$} films and {BiFeO$_{3}$}/{BiMnO$_{3}$}
  superlattices}.
\newblock \emph{\bibinfo{journal}{Sci. Rep.}}
  \textbf{\bibinfo{volume}{5}}, \bibinfo{pages}{9093} (\bibinfo{year}{2015}).

\bibitem{Diep.FCC.2014}
\bibinfo{author}{Ngo, V.~T.}, \bibinfo{author}{Hoang, D.~T.},
  \bibinfo{author}{Diep, H.~T.} \& \bibinfo{author}{Campbell, I.~A.}
\newblock \bibinfo{title}{Effect of disorder in the frustrated ising fcc
  antiferromagnet: phase diagram and stretched exponential relaxation}.
\newblock \emph{\bibinfo{journal}{Mod. Phys. Lett. B}}
  \textbf{\bibinfo{volume}{28}}, \bibinfo{pages}{1450067}
  (\bibinfo{year}{2014}).

\bibitem{Cardoso.PRB.2003}
\bibinfo{author}{Cardoso, C.~A.} \emph{et~al.}
\newblock \bibinfo{title}{Spin glass behavior in
  {RuSr$_2$Gd$_{1.5}$Ce$_{0.5}$Cu$_2$O$_{10-{\delta}}$}}.
\newblock \emph{\bibinfo{journal}{Phys. Rev. B}}
  \textbf{\bibinfo{volume}{67}}, \bibinfo{pages}{020407}
  (\bibinfo{year}{2003}).

\bibitem{Frank.CheRev.2001}
\bibinfo{author}{de~Groot, F.}
\newblock \bibinfo{title}{High-resolution {X}-ray emission and {X}-ray absorption
  spectroscopy}.
\newblock \emph{\bibinfo{journal}{Chem. Rev.}}
  \textbf{\bibinfo{volume}{101}}, \bibinfo{pages}{1779--1808}
  (\bibinfo{year}{2001}).

\bibitem{HeQing.NatComm.2011}
\bibinfo{author}{He, Q.} \emph{et~al.}
\newblock \bibinfo{title}{Electrically controllable spontaneous magnetism in
  nanoscale mixed phase multiferroics}.
\newblock \emph{\bibinfo{journal}{Nat. Commun.}}
  \textbf{\bibinfo{volume}{2}}, \bibinfo{pages}{225} (\bibinfo{year}{2011}).

\bibitem{FANG.unpublished}
\bibinfo{author}{Fang, Y.-W.}, \emph{et~al.}
\newblock \bibinfo{title}{The origin of magnetic frustration in rock-salt-type
  {Bi$_{2}$FeMnO$_{6}$}}.
\newblock \emph{\bibinfo{journal}{To be published}} .

\bibitem{Blaha.WIEN2k.2001}
\bibinfo{author}{Blaha, P.}, \bibinfo{author}{Schwarz, K.},
  \bibinfo{author}{Madsen, G.}, \bibinfo{author}{Kvasnicka, D.} \&
  \bibinfo{author}{Luitz, J.}
\newblock \emph{\bibinfo{title}{WIEN2k: An Augmented Plane Wave plus Local
  Orbitals Program for Calculating Crystal Properties}} (\bibinfo{year}{2001}).

\bibitem{Kresse.PRB.1996}
\bibinfo{author}{Kresse, G.} \& \bibinfo{author}{Furthm$\rm \ddot{u}$ller, J.}
\newblock \bibinfo{title}{Efficient iterative schemes for ab initio
  total-energy calculations using a plane-wave basis set}.
\newblock \emph{\bibinfo{journal}{Phys. Rev. B}}
  \textbf{\bibinfo{volume}{54}}, \bibinfo{pages}{11169--11186}
  (\bibinfo{year}{1996}).

\bibitem{Mont.PRL.2005}
\bibinfo{author}{Duan, C.-G.} \emph{et~al.}
\newblock \bibinfo{title}{Strain induced half-metal to semiconductor transition
  in gdn}.
\newblock \emph{\bibinfo{journal}{Phys. Rev. Lett.}}
  \textbf{\bibinfo{volume}{94}}, \bibinfo{pages}{237201}
  (\bibinfo{year}{2005}).

\end{thebibliography}


\end{document}